\documentclass[twocolumn]{aastex61}
\usepackage{graphicx}

\def\cmtres{\mbox{cm$^{-3}$}}
\def\esf{\mbox{{$\epsilon_{\rm SF}$}}}
\def\erg{\mbox{erg}}

\def\lcdm{{$\Lambda$CDM}}

\def\mg{\mbox{$M_{g,cold}$}}
\def\mpch{\mbox{$h^{-1}$Mpc}}

\def\ms{\mbox{$M_{s}$}}
\def\msun{\mbox{$M_\odot$}}
\def\msunh{\mbox{$h^{-1}$M$_\odot$}} 
\def\mv{\mbox{$M_{\rm vir}$}}

\def\rv{\mbox{$R_{\rm vir}$}}
\def\ome{\mbox{$\Omega_m$}}
\def\omel{\mbox{$\Omega_\Lambda$}}
\def\omeb{\mbox{$\Omega_b$}}

\def\re{\mbox{$R_{0.5}$}}
\def\reff{\mbox{$R_{\rm eff}$}}

\def\vmax{\mbox{$V_{\rm max}$}}

\def\tsf{\mbox{$t_{\rm sf}$}}
\def\tdep{\mbox{$t_{\rm depl}$}}
\def\tlb{\mbox{$t_{\rm lb}$}}

\def\mathnew{\mathsurround=0pt} 
\def\simov#1#2{\lower .5pt\vbox{\baselineskip0pt 
    \lineskip-.5pt\ialign{$\mathnew#1\hfil##\hfil$\crcr#2\crcr\sim\crcr}}}   
\def\simgreat{\mathrel{\mathpalette\simov >}}  
\def\'#1{\ifx#1i{\accent"13\i}\else{\accent"13#1}\fi}  

\submitjournal{ApJ}

\shorttitle{Global and radial stellar mass assembly of Milky Way-sized galaxies}
\shortauthors{Avila-Reese et al.}
\begin{document}

\title{The global and radial stellar mass assembly of Milky Way-sized galaxies}


\correspondingauthor{Vladimir Avila-Reese}
\email{avila@astro.unam.mx}

\author{Vladimir Avila-Reese}
\affil{Instituto de Astronom\'ia, Universidad Nacional Aut\'onoma de M\'exico, A.P. 70-264, 04510 CDMX, M\'exico }

\author{Alejandro Gonz\'alez-Samaniego}
\affiliation{Center for Cosmology, Department of Physics and Astronomy, University of California at Irvine, Irvine, CA 92697, USA}
\affiliation{Instituto de Radioastronom\'{\i}a y Astrof\'{\i}sica, Universidad Nacional  Aut\'onoma de M\'exico, A.P. 72-3 (Xangari),
 Morelia, Michoac\'an 58089, M\'exico}

\author{Pedro Col\'in}
\altaffiliation{Deceased}
\affiliation{Instituto de Radioastronom\'{\i}a y Astrof\'{\i}sica, Universidad Nacional  Aut\'onoma de M\'exico, A.P. 72-3 (Xangari),
 Morelia, Michoac\'an 58089, M\'exico}

\author{H\'ector Ibarra-Medel}
\affiliation{Instituto de Astronom\'ia, Universidad Nacional Aut\'onoma de M\'exico, A.P. 70-264, 04510 CDMX, M\'exico }

\author{Aldo Rodr\'iguez-Puebla}
\affiliation{Instituto de Astronom\'ia, Universidad Nacional Aut\'onoma de M\'exico, A.P. 70-264, 04510 CDMX, M\'exico }

\begin{abstract}
We study the global and radial stellar mass assembly of eight zoomed-in MW-sized galaxies produced in Hydrodynamics cosmological simulations. The disk-dominated galaxies (4)  show a fast initial stellar mass growth in the innermost parts, driven mostly by in-situ SF, but since $z\sim2-1$ the SF enters in a long-term quenching phase. The outer regions follow this trend but more gentle as more external they are. As the result, the radial stellar mass growth is highly inside-out due to both the inside-out structural growth and inside-out SF quenching. The half-mass radius evolves fast; for instance, \re($z=1$)$<0.5$\re($z=0$).  Two other runs resemble lenticular galaxies. One shows also a pronounced inside-out growth and the other one presents a nearly uniform radial mass assembly. The other two galaxies suffered late major mergers. Their normalized radial mass growth histories (MGHs) are nearly close among them but with periods of outside-in assembly during or after the mergers. For all the simulations, the archaeological radial MGHs calculated from the $z=0$ stellar-particles age distribution are similar to the current MGHs, which evidences that the mass assembly by ex-situ stars and the radial mass transport do not change significantly their radial mass distributions. Our results agree qualitatively with observational inferences from the fossil record method applied to a survey of local galaxies and from  look-back observations of progenitors of MW-sized galaxies. However, the inside-out growth mode is more pronounced and the \re\ growth is faster in simulations than in observational inferences. 

\end{abstract}


\section{Introduction}
\label{introduction}

According to the $\Lambda$ Cold Dark Matter (\lcdm) scenario, galaxies form from gas 
that cools and falls into the central regions of hierarchically growing CDM halos \citep{White+1978}. 
These halos acquire angular momentum mainly by tidal torques during the linear regime of the perturbation \citep{Peebles1969,White1984}.
The gas infall rate and consequent star formation rate (SFR) in galaxies are expected to be linked to the cosmological dark halo
mass accretion rate \citep[e.g.,][]{White+1991,vandenBosch2002,Faucher-Giguere+2011,Lilly+2013,Rodriguez-Puebla+2016}.
Within this scenario, the generic outcome of galaxy formation, under the assumption of gas angular momentum conservation, 
are disks that assemble their structure inside out \citep[e.g.,][]{Larson1976,FE1980,Gunn1982,Silk1987,Mo+1998,Avila-Reese+1998,Bouwens+2002,Somerville+2008}. 

The gas in the inside-out growing disks is transformed gradually into stars. As the result, exponential disks that follow the main observed correlations
of disk galaxies and that present negative radial age and color gradients are produced \citep[e.g.,][see for a review and more references  \citealp{Mo+2010}]{Boissier+2000,vandenBosch2000,Avila-Reese+2000,Firmani+2000, Stringer+2007,Dutton2009}.
The radial stellar and color distributions so established can be further affected by other accretion channels, like infall from 
filaments \citep{Dekel+2006} or minor mergers \citep[e.g.,][]{Barnes+1996,DiTeodoro+2014}, by radial gas flows within the disk 
\citep[e.g.,][]{Athanassoula+2003,Pezzulli+2016}, and stellar mass re-distribution by
internal dynamics processes \citep[e.g.,][]{Debattista+2006,Roskar+2008,El-Badry+2016,Berrier+2016}.
The radial flows play also a role in changing the metallicity radial distribution \citep[e.g.,][]{Bilitewski+2012}.

On the other hand, the formation of spheroids (elliptical galaxies and bulges) is believed to occur from the morphological transformation of disks, 
by mergers and internal dynamical processes \citep[see for recent reviews][]{Brooks+2016,Kormendy2016} or infall of gas with misaligned angular
momentum  \citep[e.g., ][]{Scannapieco+2009,Aumer+2014}. What does happen with the radial stellar population distribution of spheroids? 
During early gas-rich mergers, 
strong bursts of SF are expected to happen across the whole galaxy, but with higher intensity in the center due to gas dissipation and inflow. 
Late dry mergers as well as internal dynamical processes produce  gas infall to the center but also significant radial mixing of stars.  
Moreover, spheroid-dominated galaxies  across their evolution may suffer one or more phases of disk destruction and rebuilding.
All these processes involve active mixing of populations, so they contribute to produce flatter gradients.

The radial distribution of stellar populations can be also affected by processes that 
abruptly slow down the SFR, particularly in massive galaxies.  This rapid shutoff  
process of SFR is distinctive from the SF histories in the star-forming stage of galaxies, and is often referred as  {\it quenching}
\citep[e.g.,][]{Faber+2007,Martin+2007,Peng+2010}. 
Diverse mechanisms were proposed to explain the global quenching of galaxies \citep[for reviews, see e.g.,][]{Kormendy2016, Smethurst+2017} 
but less is known about the spatial behavior of quenching. According to some theoretical models, the quenching happens in the central regions 
and/or start there and then extend to the outer ones, that is, the quenching is the inside out. This could be the case for the quenching
due to negative feedback of an Active Galactic Nucleus (AGN), and due to the formation of stellar spheroids/bulges that stabilize the 
gas in the galactic disk, suppressing the SF efficiency \citep[][]{Elmegreen+2008,Martig+2009}  or due to the presence 
of bars that funnel gas to the center, where gas is exhausted by SF \citep[][]{Sheth+2005}. The last two processes are known as
``morphological'' quenching. More
recently, \citet[][see also \citealp{Tacchella+2016}]{Dekel+2014} have proposed a quenching process, based on the dissipative 
shrinkage of the high-redshift gaseous disk, the triggering of intense SF in the dense centre (blue nugget phase), and the consequent rapid
gas consumption and outflows that produce an inside-out quenching. 

Both the structural inside-out growth of galaxies and the inside-out SF quenching imply that the inner stellar populations result older than
the outer ones and that the stellar half-mass radius grows with time. An open question is which of these processes dominates in the
evolution of galaxies and what signatures evidence the dominion of one or another of these processes \citep{Lian+2017}.

\subsection{Cosmological numerical simulations}

As discussed above, the way galaxies establish their radial stellar population and mass distributions is complex 
\citep[see also ][]{Pezzulli+2016}. The involved processes are:
(1) the structural build-up from cold gas accreted from the cosmic dark matter web, (2) the in-situ SF and its 
eventual quenching across the galaxy, (3) the mass and angular momentum radial transport that gas/stars may suffer, 
and (4) the accretion of ex-situ stars and gas in mergers. 
In principle, all this complexity, which implies non-linear gravitational evolution, baryonic physics, galaxy
dynamical processes, etc., can be followed in the N-body + Hydrodynamics cosmological simulations, though 
it is important to bear in mind that the spatial resolution 
limits the ability to follow several of the multi-scale physical and dynamical processes of galaxy evolution.
Most of previous numerical works have focused on the evolution of global galaxy properties. There are only a few works that present
 a detailed study of the radial evolution of galaxies down to $z=0$  \citep[e.g., ][]{Bird+2013,Aumer+2014,Grand+2017} 
 or at high redshifts \citep[e.g.,][]{Zolotov+2015,Tacchella+2016b,Tacchella+2016}. Time is ripe to study in more detail the 
spatially-resolved evolution of simulated galaxies given that observations provide now valuable information on that, for instance, on the 
SF and stellar mass growth histories (MGHs) as a function of radius for galaxies from large surveys (see below). 

The main goal of the present paper is namely to study the radial stellar mass assembly and the establishment of the stellar age
distribution along the radius of a suite of eight simulated 'field' Milky Way(MW)-sized galaxies  presented in \citet[][]{Colin+2016}. 
The simulations do not suffer the overcooling and angular momentum catastrophe problems due to an
adequate SF/feedback implementation  \citep[see for a related discussion, e.g.,][]{Ubler+2014}. The stellar feedback effect 
in our simulations allows the formation of realistic disks. The \ms-to-\mv\ ratios of the simulated galaxies are reasonable at
all redshifts and the $z\sim 0$ properties of these galaxies are in good agreement with observations \citep[see][]{Colin+2016}. 

The MW-sized galaxies are special because they are in the peak of the stellar mass growth efficiency within CDM halos, measured
through the \ms-to-\mv\  ratio \citep[see for recent determinations e.g.,][and more references therein]{Rodriguez-Puebla+2017}. 
For less massive galaxies, this efficiency decreases likely because of the global effects of SN-driven outflows, and for more massive ones, 
likely because of the long cooling times of the shock heated gas during the virialization of massive halos and the feedback from luminous AGNs
\citep[for a recent review on all of these processes, see][]{Somerville+2015}. 
Therefore, being MW-sized galaxies less susceptible to the effects of the not-well constrained stellar and AGN feedback processes,
they are more suitable for exploring the predictions of \lcdm-based simulations/models and for comparing them with observations.

\subsection{Observational inferences}

The observational study of the radial mass growth of {\it individual} galaxies is based mainly on
fossil signals as the color, metallicity, and age gradients obtained 
from photometric, spectral energy distribution, and/or line-strength indices observations of local galaxies 
\citep[for recent works, see e.g.,][]{Wang:2011a, Pan:2015aa, Li:2015aa,Dale+2016,Kennedy+2016,Lian+2017}. A more
complete inference is 
provided by the fossil record method using Integral Field Spectroscopy (IFS) observations across the galaxies 
\citep[][]{Lin:2013aa,Perez+2013,Ibarra-Medel+2016,Gonzalez-Delgado+2016, Goddard+2017,Garcia-Benito+2017}.   
There are also detailed stellar population studies based on color-magnitude diagrams obtained for the MW and nearby galaxies
\citep[e.g.,][]{Williams+2009,Cheng+2012,Boeche+2014,Hayden+2014,Mikolaitis+2014}.
An alternative approach is to determine the instantaneous signal of the growth process, for example, 
by using the observational determination of the stellar and SFR density profiles
\citep{Munoz-Mateos+2007, Pezzulli+2015}.  The radial growth modes inferred with this approach strictly refer only to 
the current time but under some assumptions can be extrapolated to some fraction of the past history of galaxies. 

By using these different methods most of the authors have concluded that a significant fraction of the local galaxies 
formed (or are currently forming) their inner stellar masses earlier than their outskirts, 
that is, their stellar masses grow inside out. In more detail, several of these studies show that the mode and rate of radial 
mass growth may change with galaxy type, mass, environment, etc. For example, \citet{Ibarra-Medel+2016} have 
found for a large sample of galaxies from the 'Mapping Nearby Galaxies at the Apache Point Observatory' (MaNGA) survey \citep{Bundy+2015}, 
that blue/late-type galaxies follow, on average, a significantly more pronounced inside-out formation mode than red/early-type galaxies, 
with some evidence that the outer regions of the latter assembled more irregularly, likely due to minor mergers.

Under the assumption that stars form and remain in roughly the same position, the SF and mass growth histories inferred
with the fossil record methods are expected to trace the radial stellar mass assembly of galaxies. However, this assumption should be
taken with caution; a fraction of the stars could have been formed in other (secondary) 
galaxy(ies) that afterward merged with the primary one or the stellar mass could suffer radial net transport within the same primary galaxy
by dynamical processes, 
in such a way that the inferred radial MGHs could trace with some bias the real (dynamical) radial mass assembly of galaxies. It is currently a 
matter of debate how significant are these processes in galaxies, in particular, whether they produce a dramatic radial mass redistribution
or just mix at small scales the stellar populations increasing the width of the metallicity distribution with age
\citep[e.g.,][]{Sellwood+2002,Debattista+2006,Roskar+2008,Schonrich+2009,Roskar+2012,Berrier+2016}. 
Here, using our simulations, we will explore how different can be the radial archaeological MGHs from the real (current) ones
to evaluate the possible effects of radial mass redistribution by internal dynamical processes.

The radial mass assembly of galaxies has been also estimated 
from look-back time observations of a given galaxy population (for example, massive or MW-like galaxies)
and by using the stacking technique for the selected galaxy population at each redshift bin in order to get the
average evolution \citep[][]{vanDokkum+2010,vanDokkum+2013}.  
For the progenitors of MW-sized galaxies, \citet{vanDokkum+2013} have found that the mass contained inside 2 kpc grows on average 
slower than the mass outside this radius since $z\sim 2$; even more, since $z\lesssim 0.6$, the inner mass growth stops at all, while the 
outer mass continues growing. By means of this kind of direct look-back time studies, as well as semi-empirical approaches, 
it was also confirmed that the effective radius of the MW-sized progenitors increase with time but slowly, which implies 
a mild inside-out growth \citep[e.g.,][]{vanDokkum+2013,Patel+2013,Papovich+2015,Rodriguez-Puebla+2017}.

The content of this paper is as follows.
Section 2 briefly describes the simulations and the main properties of the simulated galaxies. 
Section 3 presents the half-mass radius and global/radial stellar mass growth of the galaxies, using different ways to account for the
accumulation or loss of stellar mass in the radial bins. In Section 4, we present a study of the SF regime in the innermost galaxy regions
compared to the whole galaxy in order to inquire about an inside-out quenching process that adds to the structural 
inside-out mass growth.  In Section 5, we compare our results with inferences from recent observations,
both based on the fossil record method applied to local galaxies and on look-back time observations of MW-size selected galaxy
populations. Section 6 is devoted to discuss our results and their comparisons with observations. We summarize and present our conclusions
in Section 7.  In Table 1 the acronyms and definitions
used in this paper are defined.

\begin{deluxetable}{ll}[h] 
\tablecaption{A list of acronyms and definitions used in this paper} 
\tablecolumns{2}
\tablewidth{0pc}
\tabletypesize{\scriptsize}
\label{Tacron}
\startdata
				 & \\
			         AGN & Active Galactic Nucleus\\
			         CANDELS & Cosmic Assembly Near-IR Deep Extragalactic \cr 
			          & Legacy Survey\\
			         IFS & Integral Field Spectroscopy\\
			         IMF & Initial Mass Function\\
			         \lcdm\ & Lambda Cold Dark Matter\\
			         MaNGA & Mapping Nearby Galaxies at the APO \\
				MGH & (Stellar) Mass Growth History, $\ms(<t)$\\
				\_\_\ current & Total stellar mass accumulated with time in a \cr
				& given region (Eq. \ref{MGHeq})\\
				\_\_\  in-situ & Stellar mass accumulated w/time only from particles\cr
				& formed in the given region (Eq. \ref{MGHeq} w/o  $M_{\rm *,in}$, $M_{\rm *,out}$)\\
				\_\_\ archaeo- & Cumulative stellar mass distribution as a function of  \cr
				\ \ \ logical & age from particles at $z\sim0$ in a given region\\
				MW & Milky Way\\ 
				SDSS & Sloan Digital Sky Survey\\
				SFR & Star Formation Rate\\
				sSFR & Specific Star Formation Rate\\
				SN & Supernova\\
				Sp\#D & Name of the runs, D stands for disk-dominated\\
				Sp\#L & Name of the runs, L  stands for lenticular-like\\
				Sp\#S & Name of the runs, S stands for spheroid-dominated\\
				\hline
				D/T & Disk-to-total stellar mass ratio (dyn. determined) \\
				\ms\ & Total galaxy stellar mass (within 0.1\rv) \\
				\mv, \rv & Virial mass and radius \\
				\re & Stellar half-mass radius\\
				\tdep & Depletion time (timescale at which the cold gas \cr 
				& is consumed given the current SFR) \\
				\tlb & Look-back time\\
				$<t_{\rm lb}>_{\rm mw}$ & Mass-weighted average look-back time from a\cr
				& given MGH \\
				$\Delta t_{\rm i-o}$ & Difference in $<t_{\rm lb}>_{\rm mw}$ at 0-0.5\re\ and 1-1.5\re\\ 
				\tsf &  Timescale at which the stellar mass is duplicated \cr
				&  keeping constant the current SFR (inverse of sSFR)\\
				\vmax & Maximum circular velocity\\
\enddata
\end{deluxetable}

\section{The simulations} 
\label{sec:model}

The suite of eight MW-sized simulations to be studied here were presented and discussed in 
\citet[][see for details therein]{Colin+2016}. The simulations were run using the N-body + Hydrodynamic 
Adaptive Refinement Tree (ART) code \citep[]{KKK97,Kravtsov03,Kravtsov+2005}, which incorporates a variety
of physical processes such as gas cooling, SF, stellar feedback, advection 
of metals, and a UV heating background source. The Compton heating/cooling, atomic and molecular 
cooling, and UV heating from a cosmological background radiation \citep{HM96}, are all
included in the computation of the cooling/heating rates. 
These are tabulated for a temperature range of $10^2 < T < 10^9$ K and a grid of densities,
metallicities, and redshifts using the CLOUDY code \citep[version 96b4]{Ferland98}.
Stellar particles are formed in the gas cells with $T < 9000$ K and $n_g > 1 \cmtres$, where
 $T$ and $n_g$ are the temperature and number density of the gas, respectively. 
A stellar particle of mass $m_* = \esf\times m_g$ is placed in a 
grid cell every time the above conditions are simultaneously satisfied, 
where $m_g$ is the gas mass in the cell and \esf\ is a parameter. 
We have set $\esf = 0.65$. This and the other subgrid parameters were found in \citet[][see also \citealp{Avila-Reese+2011}]{Santi+2016}
to be optimal within the context of our subgrid recipes for the  given minimal spatial resolution
attained in our simulations (109 $h^{-1}$pc) and the integration time steps. 

An ``explosive'' stellar thermal feedback recipe was used.  Each stellar particle deposits into its parent cell 
$E_{{\rm SN+Wind}} = 2 \times 10^{51}\ \erg$ of thermal energy for each star more massive 
than $M_\star= 8$ \msun\ (half of this energy is assumed to come from the type-II SNe
and half from the shocked stellar winds)
and a fraction $f_Z=$min(0.2,$0.01M_\star -0.06$) of their mass as metals.
The code also accounts for the SN Ia feedback assuming a rate that slowly increases with time and broadly peaks at 
the population age of 1 Gyr. A fraction of $1.5 \times 10^{-2}$ of mass in stars between 3 and 
8 \msun\ explodes as SNe Ia over the entire population history and each SN Ia 
ejects 1.3 \msun\ of metals into the parent cell.  For the assumed \citet{MS79} IMF ($M_\star$ in the range 0.1--100 \msun),  
a stellar particle of $10^5\ \msun$ produces 749 SNe II (instantly) and 110 SNe Ia (over several billion years). 
On the other hand, stellar particles return a fraction of their mass and metals to the surrounding (stellar mass loss).

 The thermal energy, suddenly dumped into the cell,
raises the temperature of the gas cell to values $\simgreat 10^7$ K.
Although, as a receipt to avoid overcooling \citep[see e.g.,][]{Stinson+2006},
we have disabled the radiative cooling for some time  (40 Myr)
 in those cells where young stellar particles are located, this is
almost not necessary in our simulations because for the typical densities and
temperatures found in the SF cells, immediately after the formation
of a stellar particle ($\simgreat 1\ \cmtres$
and $T \simgreat\ 10^7$ K), the cooling time is actually longer
than the crossing time \citep{DS2012}. Therefore, in most of the
cases the gas in the cell, where the newborn stellar particle is located, 
expands to the neighbor cells before radiating away its heat.

The simulations were performed for the \lcdm\ cosmology, with $\ome = 0.3$, $\omel = 0.7$, 
and $\omeb = 0.045$. The CDM power spectrum was taken from \citet{kh97} and it was normalized to 
$\sigma_8 = 0.8$, where $\sigma_8$ is the rms amplitude of mass fluctuations in 
8 \mpch\ spheres. First, eight halos of present-day masses around $10^{12}\ \msun$ 
were chosen from a low-resolution N-body dark matter-only ART simulation, run in a box
of 50 Mpc/h on a side. Except for one halo, which has a companion of comparable mass at a distance
of 0.26 \mpch, all the others are relatively isolated at $z\sim0$. Thus, the environment of the simulated
galaxies should not be associated to one of groups/clusters and can be related to what observers 
call the field. The eight selected regions were resimulated with much higher resolution and including baryons
with the N-body + Hydrodynamics ART code. The maximum level of refinement was set to
12 so that the high density regions are mostly filled, at $z=0$, 
with cells of 109 $h^{-1}$pc per side; this is the nominal spatial resolution of our simulations.
The number of dark matter particles in the high-resolution zone,
where galaxies are located, ranges from about 1.5 to 2 million and the particle mass is 
$1.02 \times 10^6\ \msunh$. 
On the other hand, the number of stellar particles in the galaxies at $z=0$ ranges from 1.2 to $5.1\times 10^5$.

\citet{Colin+2016}  have determined the kinematic spheroid-to-disk mass
ratio at $z=0$ by means of two methods. The results were close in both cases. From these results, we may classify the 
simulated galaxies as: disk dominated, highlighted with the final letter D (runs Sp3D, Sp1D, Sp7D, and Sp8D), 
lenticulars like (with spheroid-to-disk ratios not too different, highlighted with the final letter L; Sp2L and Sp6L), and spheroid dominated 
(highlighted with the final letter S; Sp4S and Sp5S). The two
latter runs are spheroid dominated because they have undergone late major mergers; during these late mergers the galaxies
had bursts of SF and at $z=0$ they are yet forming stars (5.6 and 1.4 \msun/yr, respectively);
they would correspond to the rare cases of  blue, star-forming early-type local galaxies. 
Aside from these two runs, the general evolutionary picture for the other runs is that at high redshifts ($z>1-2$) the 
disks were turbulent, with high vertical gas velocity dispersions and intense bursts of SF, but since $z\sim 1$,
the galaxies evolve quiescently, with relatively low SFRs and growing stable gaseous disks \citep{Colin+2016}. 

In \citet{Colin+2016} were presented the main properties of the eight simulated galaxies (see Table 1 therein).
In our Table \ref{properties} we reproduce some of these properties. The runs are ordered from the largest
to the lowest disk-to-total mass ratio $D/T$ at $z=0$ (column 5). The $z=0$ stellar masses range from 
$\approx 2$ to $8\times 10^{10}$ \msun, and their stellar half-mass radii,
from  2.6 to 6.8 kpc. The galaxies are only slightly above the $V_{\rm max}-\ms$ observed relation but mostly
within the intrinsic scatter around this relation. Their radii and cold gas fractions are also in rough agreement with 
observations given their masses. The rotation curves of all the galaxies are nearly flat, with  $V_{\rm max}$ 
values from 165 to 226 km/s. The stellar-to-halo mass ratios at $z\sim0$ are in agreement (within the 1$\sigma$ scatter) 
with direct and semi-empirical determinations.

\begin{deluxetable}{cccccccc}
\tablecaption{Some properties of simulated galaxies\tablenotemark{a}}
\tablecolumns{8}
\tablewidth{0pc}
\tablehead{
\colhead{(1)} &  \colhead{(2)} & \colhead{(3)} & \colhead{(4)} & \colhead{(5)} & \colhead{(6)} & \colhead{(7)} \\
\colhead{Run} & \colhead{\mv} & 
\colhead{\ms} & \colhead{\re} & \colhead{D/T}
& \colhead{``Morph''} & \colhead{$\Delta t_{\rm i-o}$} \\
 & $10^{12}$\msun & $10^{10}$\msun & kpc &  &  & (Gyr)  }
\startdata
Sp3D & 0.99 & 5.3 & 6.4 & 0.81 & disk-dom & 3.29 \\
Sp1D & 0.84 & 1.8 & 6.0 & 0.75 &  disk-dom & 2.06  \\
Sp7D & 1.09 & 5.4 & 5.2 & 0.69 &  disk-dom & 2.04  \\
Sp8D & 1.20 & 6.3 & 5.9 & 0.64 &  disk-dom & 3.37 \\
Sp2L & 0.83 & 3.3 & 4.9 & 0.43 &  lentic. & 1.18 \\
Sp6L & 0.97 & 2.3 & 2.6 & 0.30 &  lentic. & 2.53 \\
Sp4S & 1.56 & 8.4 & 6.8 & 0.25 &  sph-dom & 0.80 \\
Sp5S & 1.05 & 4.1 & 3.3 & 0.09 &  sph-dom & 0.02 \\
\enddata
\tabletypesize{\scriptsize}
\tablenotetext{a}{See definitions in Table 1.} 
\label{properties}
\end{deluxetable} 

\section{The structural assembly and stellar mass growth histories} 
\label{MGHs}

\subsection{Evolution of the stellar half-mass radius}
\label{R-evolution}

Figure \ref{R-evolution} presents the  evolution of the stellar half-mass radius, $\re$, for the eight MW-sized simulations: 
upper panel for the disk-dominated galaxies and lower for the spheroid-dominated ones. The radii were normalized to 
their $z=0$ values. In all the cases \re\ grows rapidly with time.  For example, for the disk-dominated
 galaxies, \re\ at $z=1$ is $\approx 0.3-0.5$ \re\ at $z=0$. As mentioned in the Introduction, the structural 
growth of disks formed from gas that initially follows the \lcdm\ halo mass and angular momentum distributions, and that 
cools and settles down in centrifugal equilibrium conserving angular momentum, happens the inside out. 
The thick gray dashed line in figure \ref{R-evolution} corresponds to such kind of predictions for a 
$\approx 10^{12}$ \msun\ present-day halo \citep[][see also e.g., \citealp{Somerville+2009}]{Firmani+2009};  
in this model, the SF in the disk is triggered by the Toomre gas gravitational instability criterion and selfÐregulated 
by a balance between the energy input due to SNe and the turbulent energy dissipation in the ISM; the Kennicutt-Schmidt 
relation is naturally recovered by the model. 

\begin{figure}[htb!]
\includegraphics[width=\columnwidth]{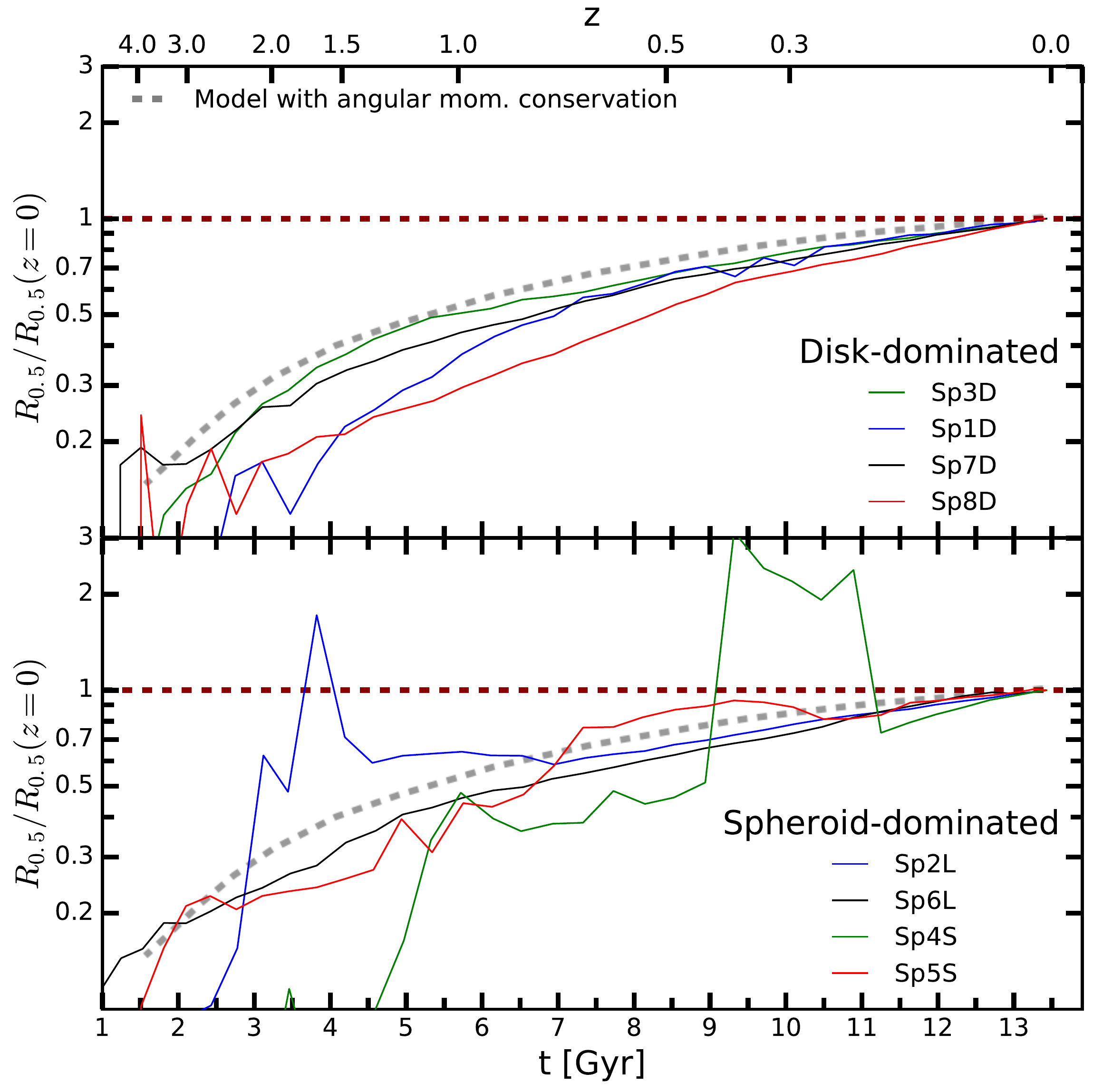}
\caption{Normalized half-mass radius evolution of the disk-dominated (upper panel) and spheroid-dominated (lower panel) galaxies. The 
normalization is with respect to the \re\ values at $z=0$. The gray line shows the \re\ evolution from a \lcdm--based galaxy evolutionary model
(detailed angular momentum conservation is assumed) corresponding to a present-day halo of $10^{12}$ \msun\  \citep{Firmani+2009}. }
\label{R-evolution}
\end{figure}

Our numerical results for the disk-dominated galaxies, show a faster $\re$ evolution up to $z=1-2$ than the predicted 
for the simple \lcdm\--based galaxy evolutionary models. While this could be by many reasons, a possibility is that in simulations, apart from 
the structural inside-out growth of stellar mass,  an inside-out quenching of SF happens. 
In this case, the inner regions slow down or totally quench the SF first, braking
their stellar mass growth, while the outer regions keep forming stars, in such a way that their stellar masses grow now
faster than in the inner regions; as the result, \re\ increases more rapid with time. We will see below that this indeed happens in most of the runs.  
For the lenticular-like and spheroid-dominated galaxies, \re\ has periods of increase and decrease related to the mergers, 
excepting run Sp6L that does not suffer mergers since early epochs.

\subsection{Global and radial stellar mass growth histories}
\label{results}

Our main goal is to study the global and radial stellar mass growth of the simulated MW-sized galaxies in the context
of the \lcdm\ cosmology. We dissect the galaxies at any time in cylindrical bins with radii 
($0,0.5$), ($0.5,1)$, ($1,1.5)$, and ($1.5,2$) \re(0),\footnote{\re(0) is
the present-day stellar half-mass radius. We have chosen these radial bins in order to compare them 
with observational inferences (Section \ref{comparisons}).} and height $\pm 2$ kpc 
above the disk plane. Then, we measure within these fixed bins the accumulation of stellar mass in three different ways:

 {\it (1) Current (real) mass growth histories (MGHs)}, defined as the stellar mass accumulated at each snapshot 
(redshift) within the radial bins. These histories account for the amount of mass in stellar particles added or subtracted 
at each radial bin between two snapshots (epochs): 
\begin{equation}
\Delta M_*  = M_{\rm *,in-situ} + M_{\rm *,in}  - M_{\rm *,out} - M_{\rm *,ml},
\label{MGHeq}
\end{equation}
where 
$M_{\rm *,in-situ}$ is the mass in stellar particles formed in the bin (in situ), 
$M_{\rm *,in}$ is the mass in stellar particles that came from other radial bins or from outside the galaxy 
(by mergers, for instance), $M_{\rm *,out}$ is the stellar mass in particles that left the radial bin, and 
$M_{\rm *,ml}$ is the mass loss by winds suffered by the stellar particles in the bin depending on their formation epochs 
and the current time.

{\it (2) In-situ stellar MGHs}, defined as the stellar mass accumulated by stellar particles formed only in situ at each
radial bin, not taking into account gained or lost stellar particles from/to other regions (but taking into account the stellar mass loss of the in-situ formed
stellar particles). This is equivalent to Eq. (\ref{MGHeq}) without the terms $M_{\rm *,in}$ and $M_{\rm *,out}$. 
Note that the in-situ MGH is related to the {\it cumulative SF history} typically used
in the literature.

{\it (3) Archaeological MGHs}, constructed from the information of the stellar particles in the last snapshot ($z=0$), that is, in this
case we do not use the historical information contained in previos snapshots. 
For each stellar particle at $z=0$, we know its formation time (age), its initial mass, and the account of mass lost by winds 
as a function of time. Therefore, from the distribution of stellar mass fractions as a function of age (look-back time) in a given 
region of the $z=0$ galaxy we can calculate the accumulated stellar mass at each look-back time in that region.
This is conceptually similar to the observational determinations obtained by means of the fossil record method 
\citep[see e.g.,][]{Perez+2013,Ibarra-Medel+2016, Goddard+2017}; since in this case the information comes only from 
the observed stellar populations, it is not immediate to know whether stars were formed at the ``observed'' 
position or in other regions.

Following, we present the three types of MGHs described above for each one of our eight MW-sized galaxy simulations. 
The radial MGHs are {\it normalized} to their corresponding masses at $z=0$ in such a way that the relative growth rates among 
different radial bins and different galaxies can be compared. In this sense, the figures show rather than the absolute mass contribution
of each radial bin, {\it the rate at which the mass grows in each bin}. The panels in the figures to be shown below are ordered
from the most to the least disk-dominated galaxies at $z=0$.

\begin{figure*}[htb!]
\includegraphics[width=\textwidth]{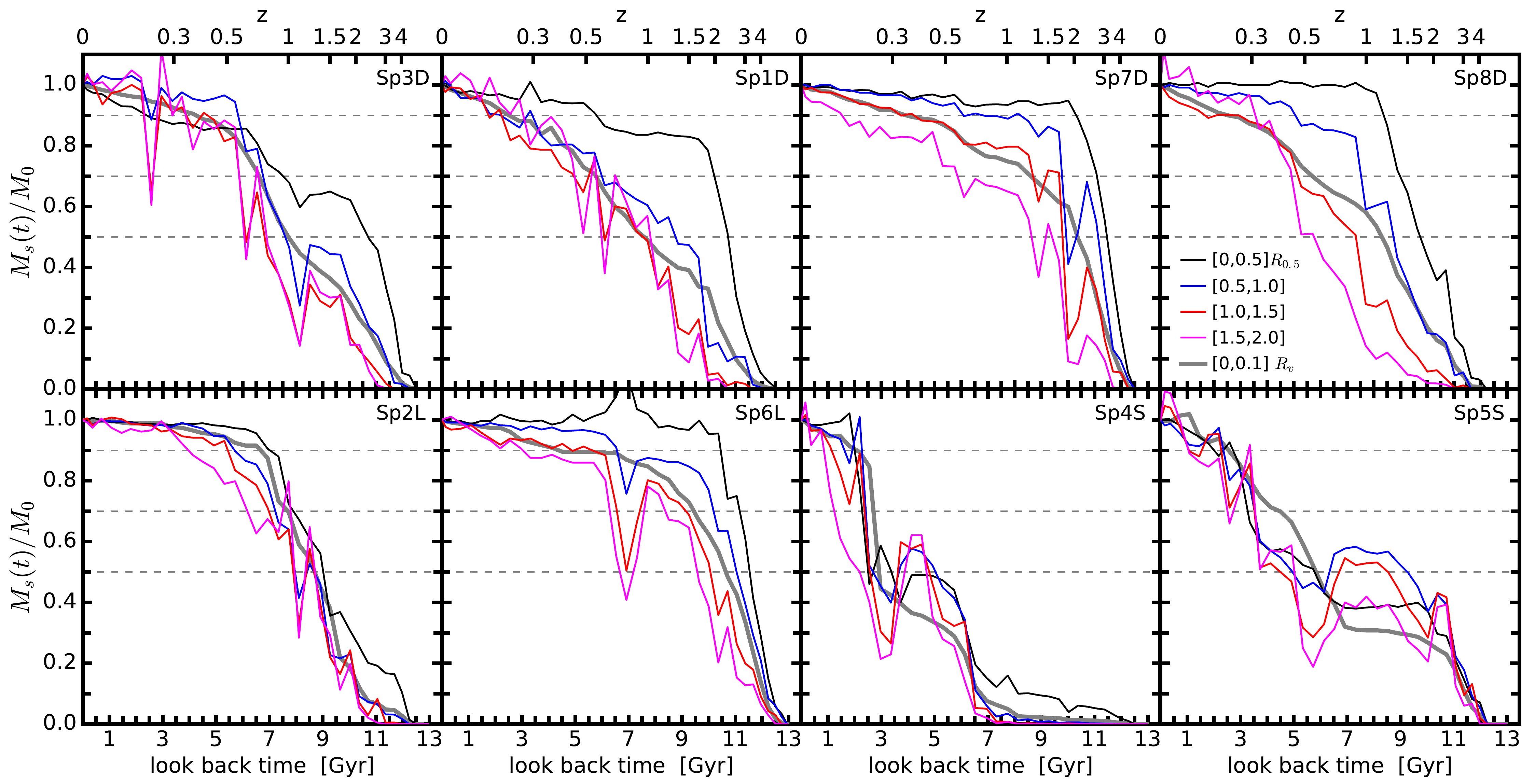}
\caption{Global (gray line) and radial (colored lines) current MGHs for the eight simulated galaxies. The color code of each radial 
bin is plotted in the inset of the upper right panel. Each MGH is normalized to the present-day stellar mass contained in the respective
radial bin or in the whole galaxy.  The radial bins are defined in terms of \re\ as measured at $z=0$ and they remain the same in 
physical unities at all redshifts.  The current MGHs account for the stellar mass directly measured at each epoch.
From left to right and from top down the kinematical disk-to-total mass ratio of the galaxies decreases. }
\label{currentMGHs}
\end{figure*}

{\bf Current stellar MGHs:} 
Figure \ref{currentMGHs} presents these MGHs for each one of the radial bins defined above (solid lines)
as well as for the total galaxy (thick gray lines).  In all the cases, the inner 
radial bins accumulate most of the time earlier their stellar masses than the outer bins, that is, the mass growth
is the inside out. The most pronounced inside-out trends are for the disk-dominated galaxies, specially Sp7D and Sp8D, while the  
least pronounced trends are for the spheroid-dominated Sp4S and Sp5S galaxies, which suffered late major mergers. 
At some epochs, the mass fractions in a given radial bin can decrease. This is most evident in the Sp4S and Sp5S galaxies, 
at epochs before or during their major stellar mergers; strong radial mass redistribution processes take place during this phase and finally
the mass increases at all radii significantly due to both the addition of stellar particles from the secondary galaxy and to enhanced in-situ star 
formation (see below).

The horizontal dotted lines in figure \ref{currentMGHs} indicate when each radial bin, or all the galaxy, has attained 50, 70, and 90 per cent 
of the respective $z=0$ 
stellar mass.  The assembly time differences between the innermost radial bin (0--0.5\re) and the outer bin (1--1.5\re) for the corresponding
mass fractions of 70 per cent, $\Delta t_{\rm in-out}$(70\%), are 1.63, 4.40, 2.40, and 3.84 Gyr for the disk-dominated Sp3D, Sp1D, Sp7D, and Sp8D galaxies, respectively, showing a clear inside-out behavior. These assembly time differences are lower for the spheroid-dominated 
galaxies, excepting for Sp6L, which shows a pronounced inside-out mass assembly. 

For all the simulated galaxies, but Sp4S and Sp5S, the innermost regions ($0-0.5$\re) suddenly slow down or cease 
at all their stellar mass growth since early look-back times that range from $\approx 10$ Gyr (Sp7D and Sp6L) to 
$\approx 6$ Gyr (Sp2L, Sp3D). This change in the stellar mass growth velocity is also observed in the outer regions but it
is typically more gentle and delayed to later epochs than in the innermost regions. This behavior could be 
due to preferential late stellar mass accretion (mergers) in the outer regions or to significant radial mass redistribution from inside out. 
 However, as we will see below, the main reason is that galaxies, besides the structural inside-out growth, suffer an inside-out SF quenching. 

\begin{figure*}[htb!]
\includegraphics[width=\textwidth]{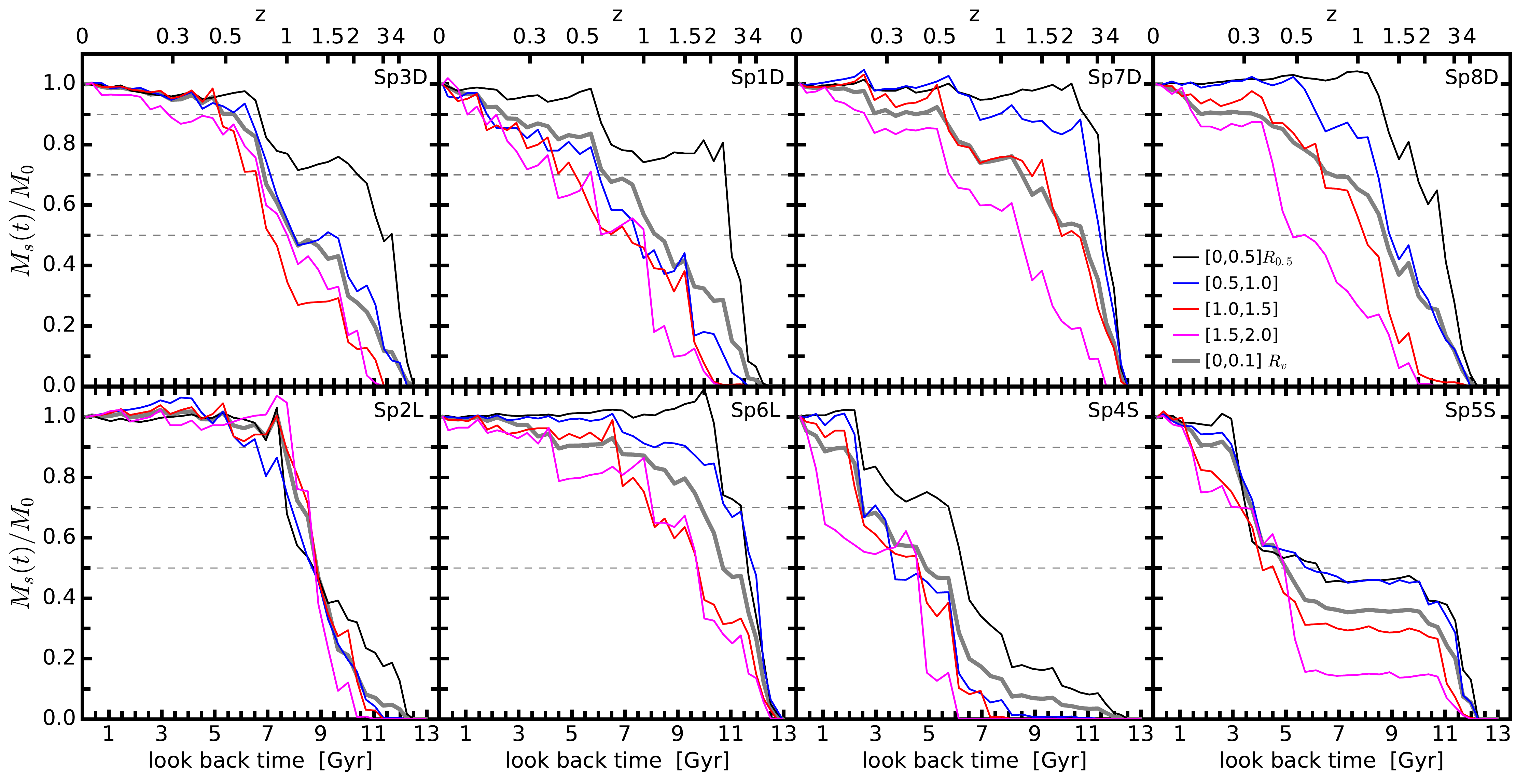}
\caption{As in figure \ref{currentMGHs} but for the in-situ MGHs. The in-situ MGHs account only for the mass accumulated in stellar particles formed
in the galaxy or in the given radial bin, taking into account the stellar mass loss by winds but not the loss or gain of ex-situ stellar particles.  }
\label{insituMGHs}
\end{figure*}

\begin{figure*}[htb!]
\includegraphics[width=\textwidth]{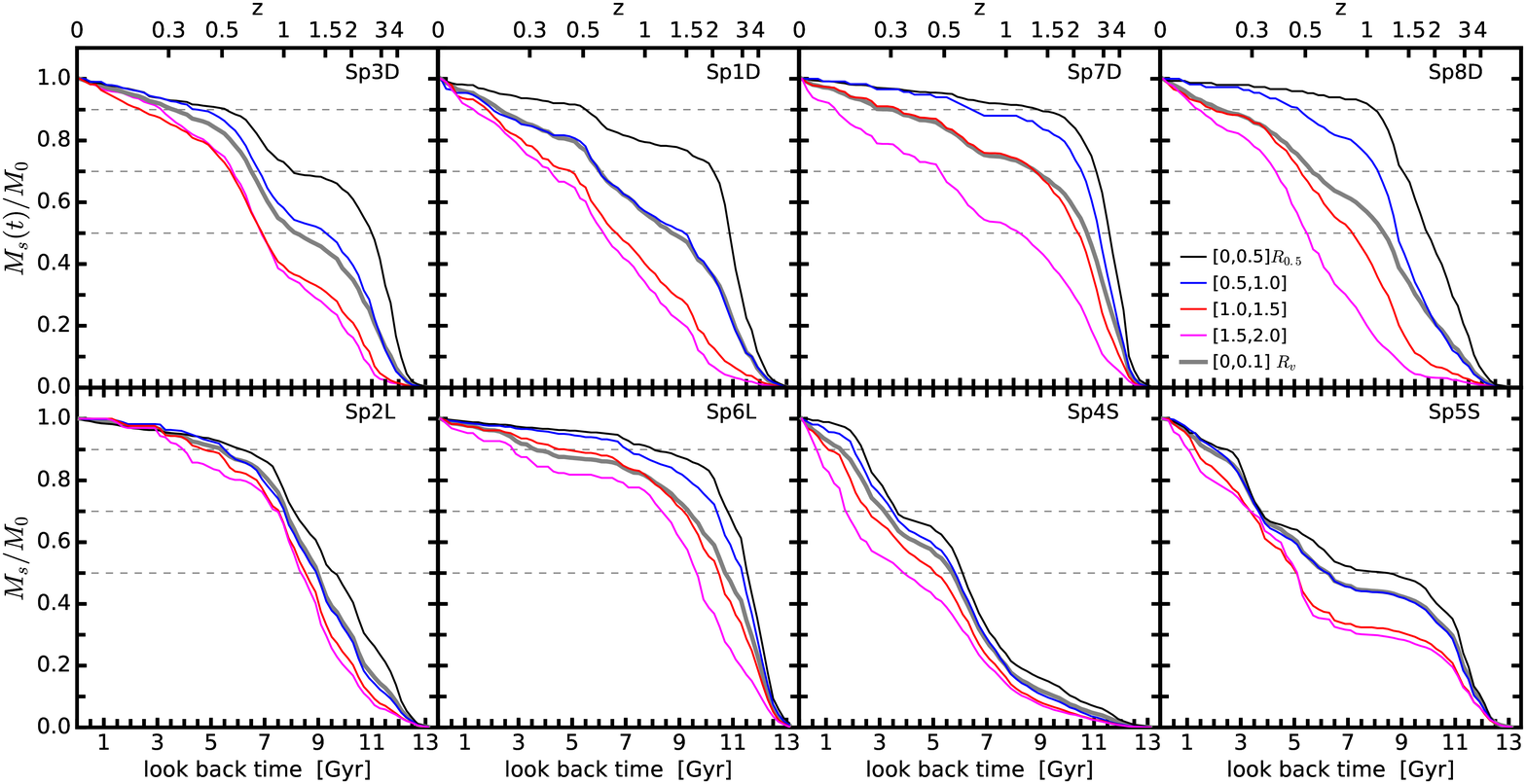}
\caption{As in figure \ref{currentMGHs} but for the archaeological MGHs. The archaeological MGHs are constructed from the age distributions
of the stellar particles measured at $z=0$. These MGHs can be  compared to the fossil record inferences from IFS observations.  }
\label{archMGHs}
\end{figure*}

{\bf In-situ stellar MGHs:} 
Figure \ref{insituMGHs} is as figure \ref{currentMGHs} but for the in-situ MGHs or cumulative SF histories. 

By comparing the in-situ and current radial normalized MGHs we find little differences, which shows
that {\it the radial stellar mass growth of the simulated galaxies is mainly driven by in-situ SF}, 
without significant contribution of mergers and/or large-scale radial mass flows.
In more detail, the current inner normalized MGHs in most of the cases are slightly shifted to lower fractions at a given time than the corresponding 
in-situ MGHs. This is mainly because, apart from the growth by in-situ SF, the (inner) stellar mass assembles by some accretion events (minor mergers,
which in spiraling motion tend to fall to the galaxy center). 
The normalized current MGH shifts to a lower mass fraction with respect to the normalized in-situ MGH each time a merger happens. Some
(little) large-scale radial mass flows are also possible, contributing this to make slightly different the current and in-situ MGHs.

The in-situ normalized MGHs show that SF and its quenching in the simulated galaxies proceeds inside out, even in the Sp4S 
and Sp5S runs (where late major mergers happened): the in-situ cumulative SF at early epochs grows faster in the inner regions than 
in the outer ones, but then, the in-situ SF suddenly slowdowns or even quenches completely in the innermost (0--0.5 \re) regions while 
the outer ones keep yet increasing their masses due to in-situ SF. The only exception is run Sp2L, which shows a nearly radially 
homogeneous mass growth by in-situ SF, with periods of even outside-in SF. As it will be seen in Section \ref{SFHs},
the SF in this galaxy becomes very inefficient in its outer regions.  

In Appendix \ref{mass-distribution}, we present complementary evidence of the strong inside-out growth of the simulated galaxies: 
we show the evolution of the radial cumulative mass distribution of in-situ formed stars, and the comparison of the measured total
radial stellar mass distributions with the ones of the in-situ formed stars. Further, in Appendix \ref{migration} we explore the net radial
transport of stellar particles (only for the Sp8D and Sp6L runs) and find that they may displace on average up to 
1-2 kiloparsecs away (1$\sigma$) from their birth place but roughly in both radial directions, not causing such a significant 
global mass redistribution.

{\bf Archaeological stellar MGHs:} 
Figure \ref{archMGHs} is as figures \ref{currentMGHs} and  \ref{insituMGHs} but for the archaeological stellar MGHs constructed from the age distributions
of the stellar particles measured at $z=0$. In general, these radial MGHs are very similar to the current ones, which shows that 
neither mergers nor global radial stellar transport affect too much the radial structural evolution of our \lcdm\ simulated MW-sized galaxies.
Only during the major mergers in the Sp4S and Sp5S runs the normalized archaeological radial MGHs differ significantly from the current ones. The 
archaeological MGHs always increase while the current MGHs may decrease and then suddenly increase due to radial mass redistribution
and mass accretion during galaxy interactions and mergers. 
The archaeological radial MGHs can be compared to the fossil record inferences from IFS observations (see Section \ref{comparisons}). 

\begin{figure}[htb!]
\includegraphics[width=\columnwidth]{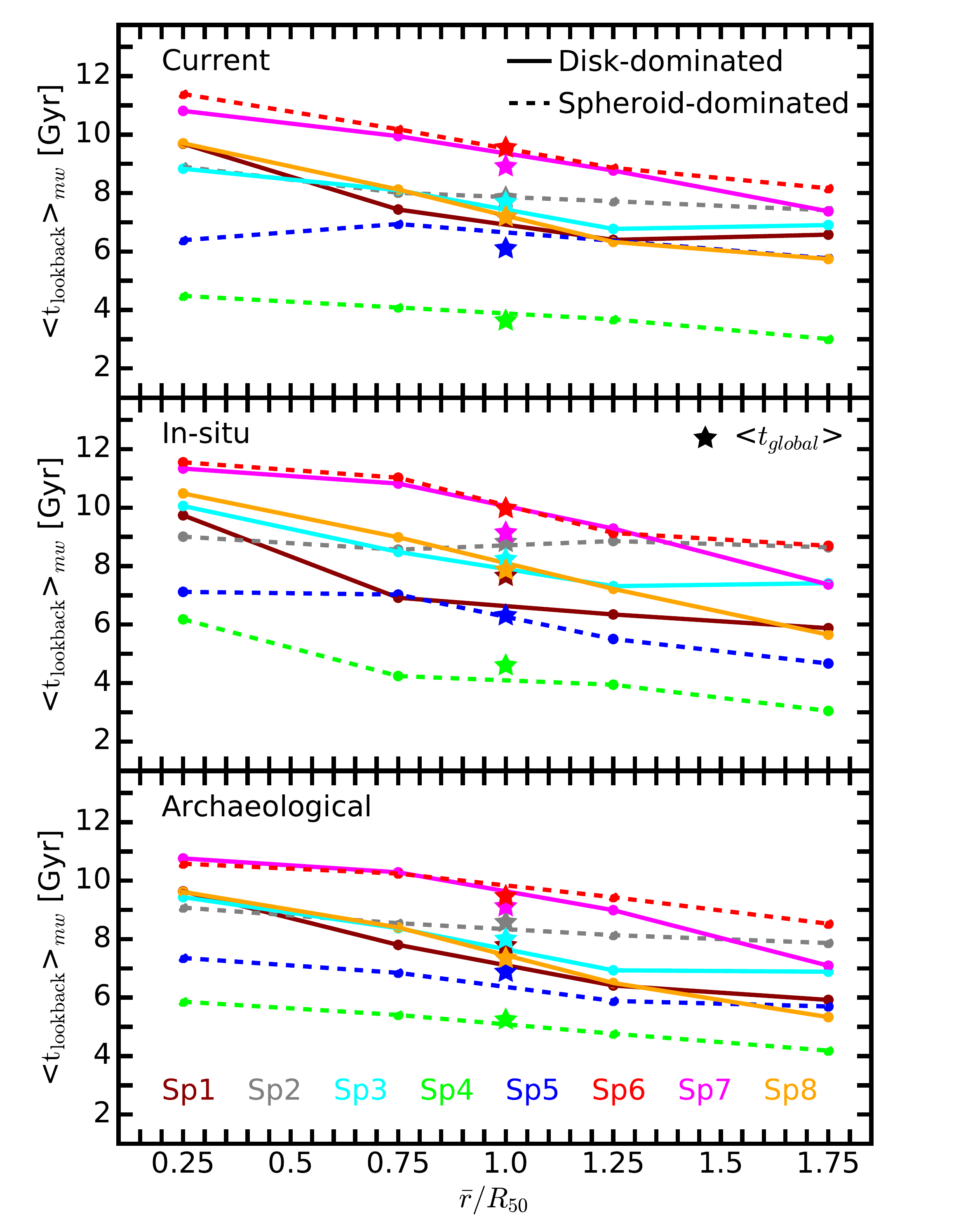} 
\caption{Average mass-weighted look-back times of each one of the radial bins used in figures \ref{currentMGHs}--\ref{archMGHs}
(solid lines and dashed lines are for the disk- and spheroid-dominated galaxies, respectively). From top to bottom, the ``age'' profiles
correspond to the current, in-situ, and archaeological MGHs, respectively. The stars indicate the corresponding global average 
mass-weighted look-back times.
}
\label{age-profile}
\end{figure}

\subsection{Quantifying the radial stellar mass assembly}

The different radial MGHs plotted in figures \ref{currentMGHs}--\ref{archMGHs} show that the simulated
MW-sized galaxies assemble (driven mostly by in-situ SF) and slow down/quench their SF
the inside out. To quantify and compare the radial differences in the normalized stellar MGHs of galaxies, 
we calculate the average look-back time of each one of the radial MGHs, $<t_{\rm lb}>_{\rm mw}$,
that is, we collapse the given cumulative mass track into an unique quantity. 
This is calculated as the sum of each snapshot look-back time weighted by the fraction of the stellar mass increased/decreased 
since the previous snapshot. In figure \ref{age-profile}, for each simulated galaxy, we plot $<t_{\rm lb}>_{\rm mw}$ for the four radial 
bins used in previous subsections. From top down, the panels are for the current, in-situ,  and archaeological MGHs. 
Note that in the last case, $<t_{\rm lb}>_{\rm mw}$ is related to
the mean mass-weighted age of the stars in the given radial bin. Solid (dashed) lines are for the disk- (spheroid-)dominated galaxies. 
The current, in-situ, and archaeological mass-weighted ages of the whole galaxy are shown for each run with a star in the corresponding 
panels of figure  \ref{age-profile}.

The average mass-weighted look-back times decrease in general with radius: the stellar mass assembly and its braking happened  
earlier in the inner regions than in the outer ones. Such inside-out trends are similar for the current, in-situ, 
and archaeological MGHs. In general, the average times are slightly higher (older ages) for the in-situ MGHs than for the current MGHs. As
discussed above, this is mainly because the current MGHs, besides the stellar particles formed in situ, take into account the accretion of 
ex-situ particles (some of them coming from relatively late mergers);  that is, the current MGHs refer both to the intrinsic SF 
and the (dynamical) stellar mass assembly in the given region. The main differences between the in-situ and current average times 
are namely for runs Sp4S and Sp5S, which suffer late major merges.

\begin{figure*}[htb!]
\includegraphics[width=\textwidth]{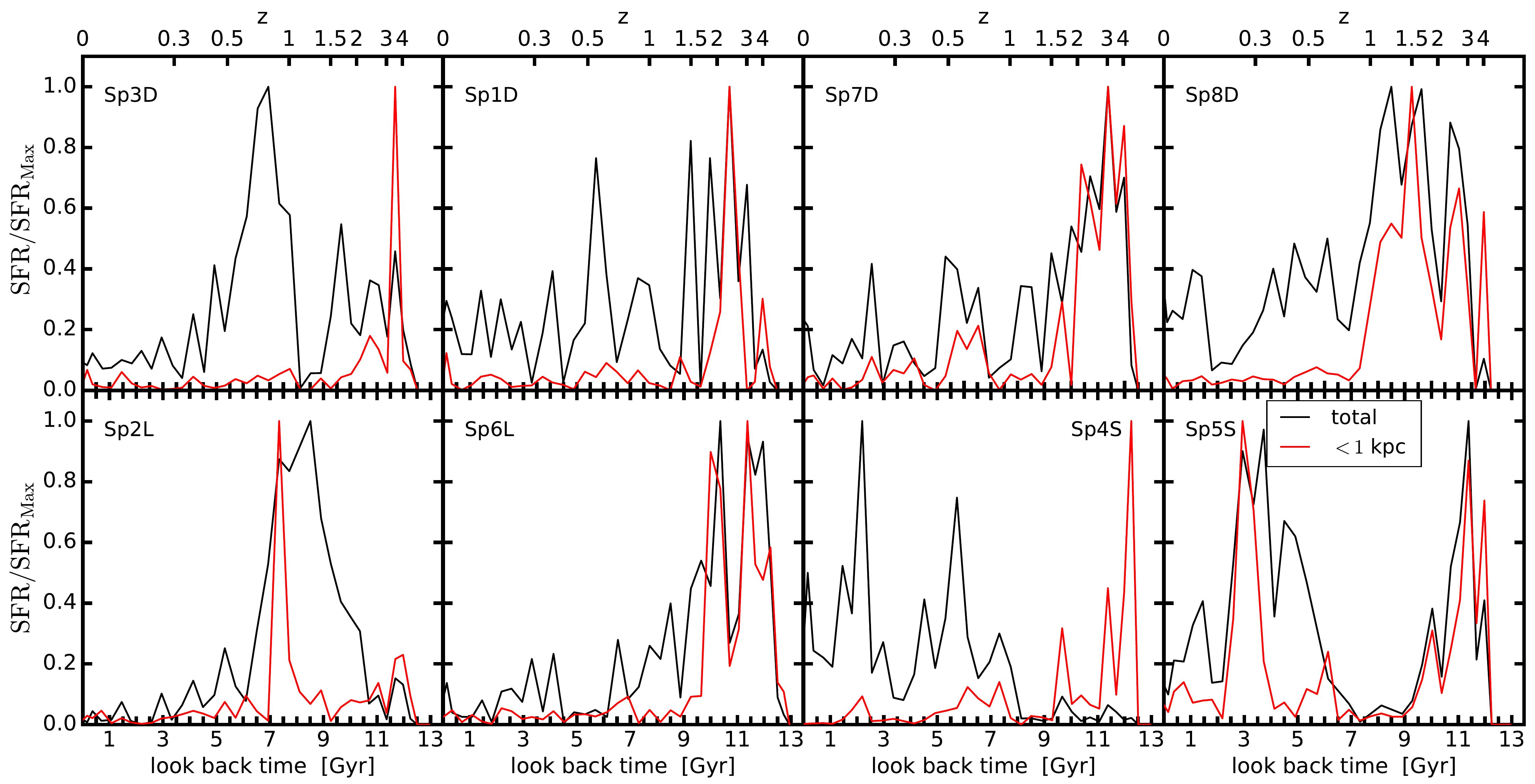}
\caption{ Evolution of the global SFR (black lines) and the SFR inside 1 kpc (red lines) of the simulated galaxies. The SFR's are
normalized to their corresponding maximum values. The SFR histories of the central regions tend to be qualitatively different 
to those of the whole galaxy.}
\label{SFRH}
\end{figure*}

The disk-dominated galaxies have the steepest $<t_{\rm lb}>_{\rm mw}$ profiles. 
For the current MGHs, the average mean look-back times of the $0-0.5$\re\ radial bin range from $\approx 8.8$ to 10.8 Gyr, 
while these times for the $1-1.5$\re\ radial bin range from $\approx 6.4$ to 8.8 Gyr. The time differences 
for each galaxy, $\Delta t_{\rm i-o}$, range from $\approx 2$ to 3.4 Gyr (see column 7 of Table \ref{properties}); 
these time differences in the case of the in-situ and archaeological MGHs are very similar. 
 
Interesting enough, the run Sp6L, while spheroid-dominated, also shows a pronounced inside-out formation 
($\Delta t_{\rm i-o}= 2.5$ Gyr); 
this is actually the earliest formed galaxy among all the runs and its large spheroid is probably the result of long disk secular 
evolution since this galaxy did not suffer 
significant mergers accross its evolution.  On the other hand, the runs Sp4S and Sp5S, which are spheroid-dominated and suffered recent 
major mergers, show small time differences among the different radial bins; these galaxies are the latest assembled in our suite 
of simulations. The run Sp2L is the one with the flattest  $<t_{\rm lb}>$ profiles; for the in-situ profile, the outer regions are even 
slightly older than the inner ones. 

 The mean look-back times of the radial archaeological MGHs are in most of the cases slightly older (typically by $<0.5$ Gyr) than those of 
 the current MGHs. The largest differences are for the run Sp4S, up to $1.5$ Gyr at all radii. Since this galaxy has suffered a recent
 major merger, its stellar mass assembly (with contribution of accreted ex-situ stars) is younger on average than the ages of the stars that 
 compose this galaxy at $z=0$. So, in the cases where galaxies suffered late major mergers, the archaeologically inferred
 MGHs and average mass-weighted ages constrain their stellar mass assembly histories with some shift to earlier epochs with respect to 
 the real ones.

\section{Central/global star formation and gas fraction histories}
\label{SFHs}

The radial stellar mass growth of our simulated galaxies is driven by in-situ SF. According to the radial in-situ MGHs presented in the previous Section,
the inner regions tend to halt their mass growth at high redshifts. Following, we explore in more detail the SFR\footnote{The SFR is measured 
as the mass sum of all the stars younger than 100 Myr at a given epoch divided by this time. We have experimented
with other times (from 40 to 200 Myr) and the results are very similar in all the cases. } 
history of the central 1 kpc region compared to the corresponding SFR history of the whole galaxy.
Figure \ref{SFRH} displays these histories for the eight runs; black and red lines are for the central 1 kpc and the whole galaxy histories, respectively. 
To compare the shapes of both SFR histories, we have normalized them to their corresponding maximum values (the absolute SFR values within 
the small 1 kpc region are much lower than those of the whole galaxy). For most of the runs, the central 1 kpc region forms stars actively at early
epochs (with fluctuations),  and then the SFR strongly and abruptly decreases (quenches). Instead, the SFR histories of the whole
galaxies present a less abrupt phase of final decreasing and show even some significant SF `bursts' at later epochs 
(see for example, runs Sp3D, Sp1D and Sp8D). So, the SFR histories of the innermost regions are 
qualitatively different to those of the whole galaxy, in particular due to the early quenching of SF in the innermost regions.  
For runs Sp4S and Sp5S, the situation is slightly different since these galaxies suffered major mergers after $z\sim 1$. 

We have measured also how the cold ($T\le1.5\times 10^4$ K) gas-to-stellar mass ratio, \mg/\ms, changes with time both for the 
whole galaxy and within the central 1 kpc. This ratio is related to the current gas reservoir and SF efficiency with respect
to the past SF. By dividing both the numerator and denominator by the SFR, we have that \mg/\ms = \tdep/\tsf, where 
$\tdep \equiv\mg$/SFR is the depletion or gas consumption time, and $\tsf\equiv \ms/$SFR is the SF timescale at which 
the stellar mass duplicates if keeping constant the current SFR (\tsf\ is the inverse of the specific SFR, sSFR). When  
$\tdep/\tsf\sim 1$ or higher,  there is a large gas reservoir and/or the SF process is very efficient. When $\tdep/\tsf<<1$, 
the gas reservoir is poor and/or the SF process becomes very inefficient as compared to the average past one.  
For example, for their high-redshift simulations of massive galaxies, \citet{Tacchella+2016b} find that when the gas
replenishment time becomes larger than $10\times \tdep$, which roughly corresponds to $\tsf\simgreat 11\times \tdep$, 
then the SF process enters in a long-term quenching phase. 

\begin{figure}[htb!]
\includegraphics[width=\columnwidth, height=5.5cm]{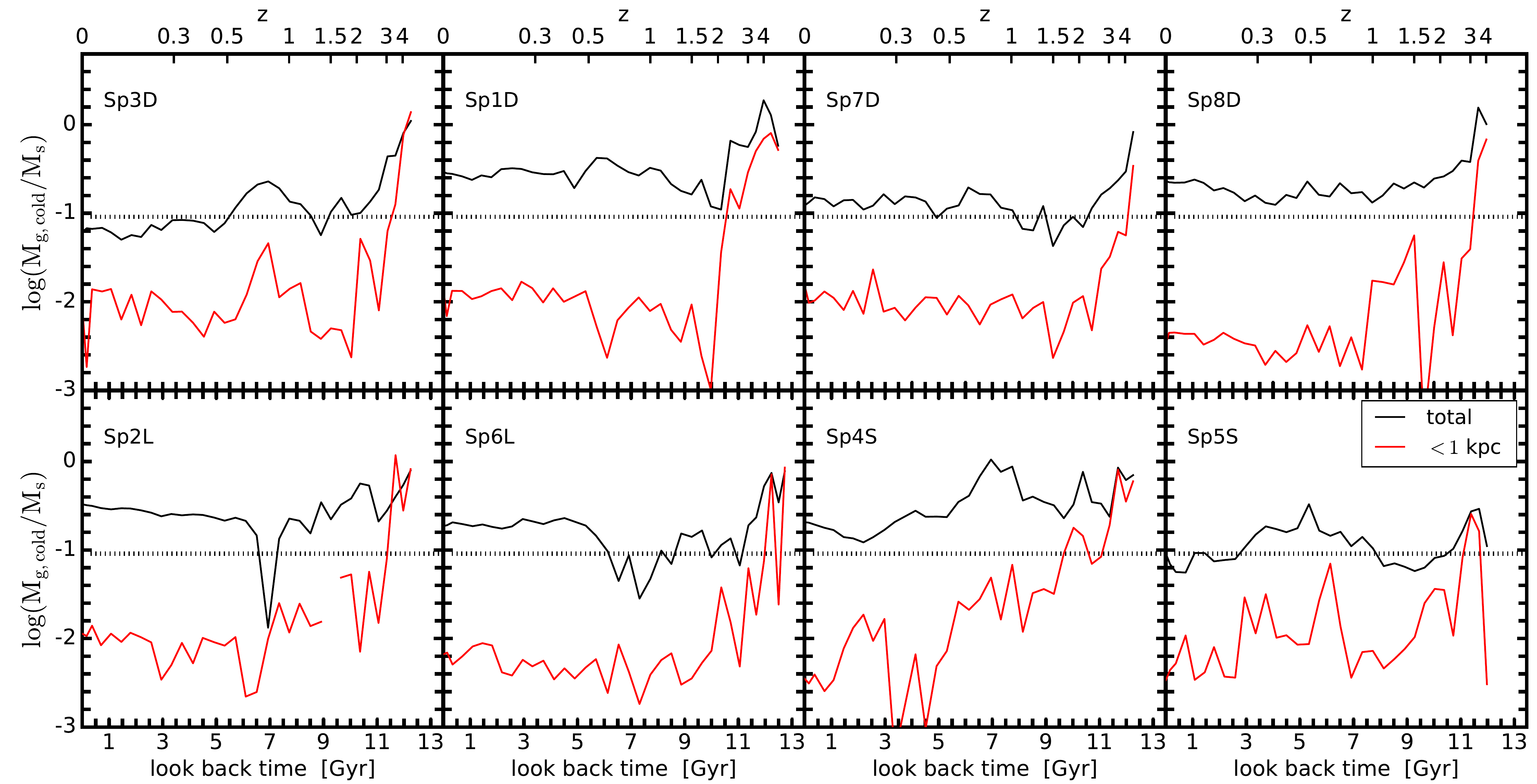}
\caption{Evolution of the cold gas-to-stellar mass ratio (\mg/\ms=\tdep/\tsf) inside 1 kpc (red lines) and
for the whole galaxy (black lines). The horizontal dotted line indicate a value of \tdep/\tsf=1/11. Below this value, the SF is expected to 
enter in a long-term quenching phase, see text. }
\label{Rgas}
\end{figure}

\begin{figure}[htb!]
\includegraphics[width=\columnwidth, height=5.5cm]{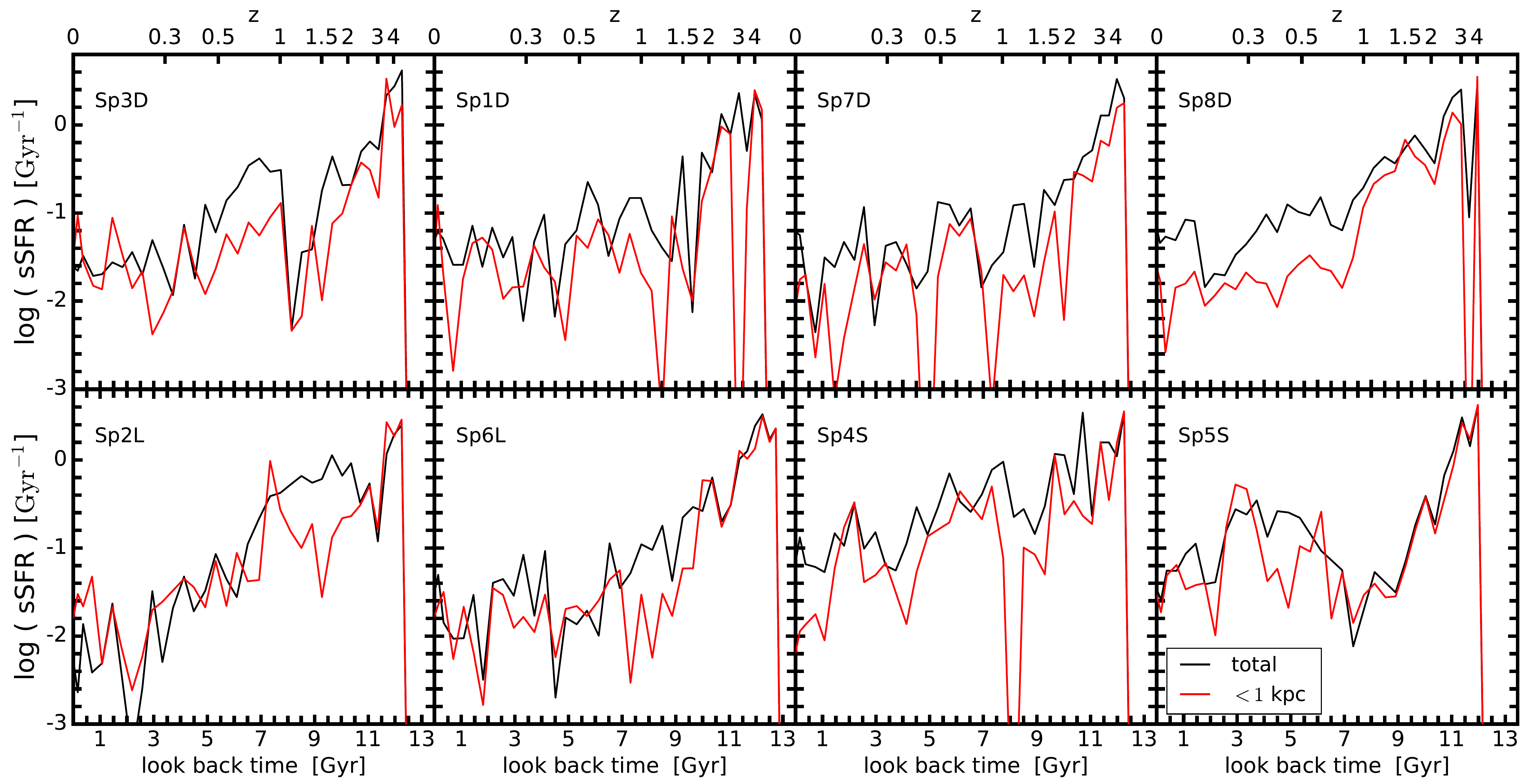}
\caption{Evolution of the global sSFR (black lines) and the sSFR inside 1 kpc (red lines).  }
\label{sSFR}
\end{figure}

\begin{figure}[htb!]
\includegraphics[width=\columnwidth, height=5.5cm]{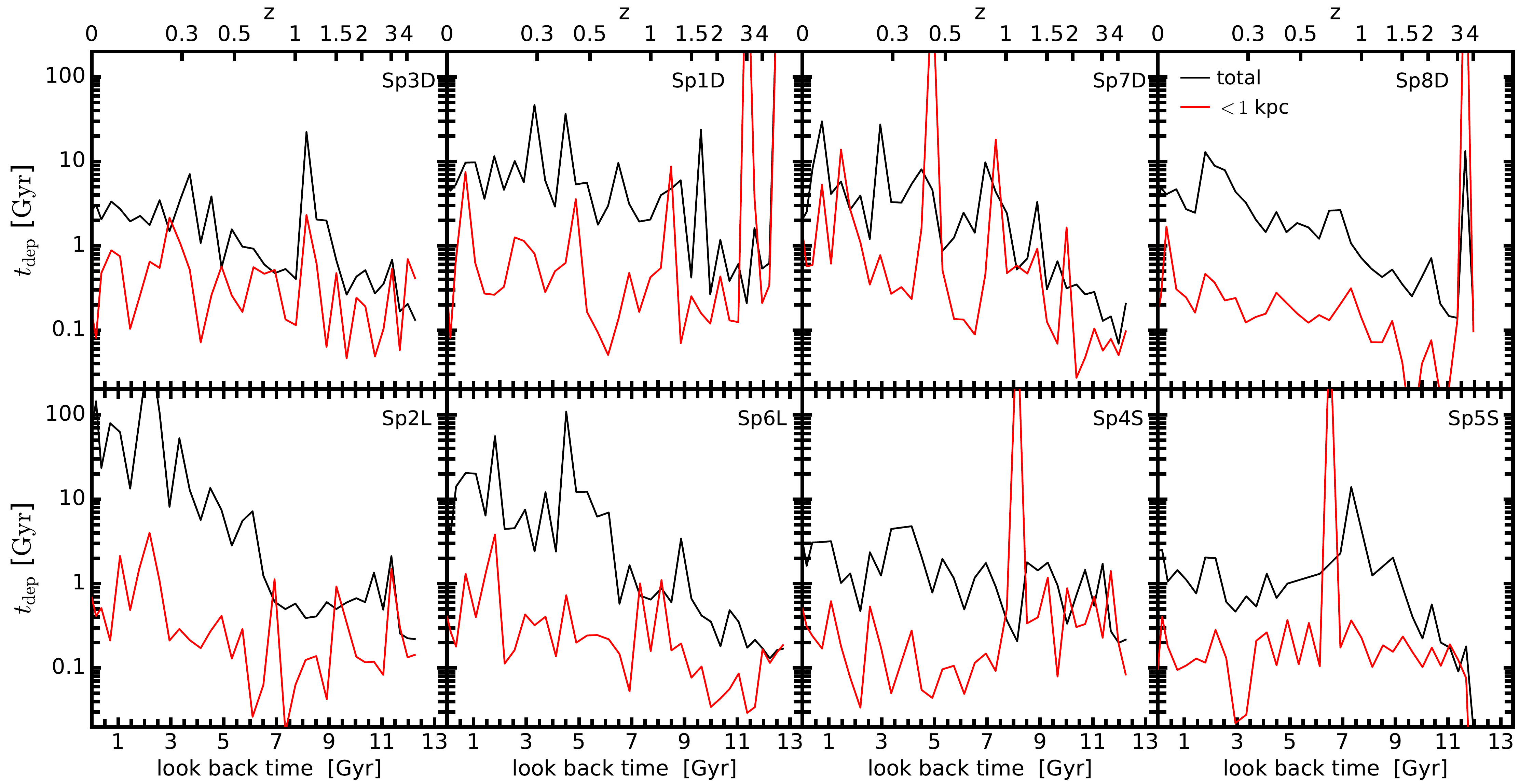}
\caption{Evolution of the global depletion time (black lines) and the depletion time inside 1 kpc (red lines). }
\label{t-depletion}
\end{figure}

Figure \ref{Rgas} shows the evolution of \mg/\ms\ (= \tdep/\tsf) within the central 1 kpc (red line) and for all the galaxy (black line), 
for our 8 simulated MW-sized galaxies. At early times ($z\sim 4$), $\mg/\ms\sim 1$ for both the 
inner 1 kpc and the whole galaxy. Galaxies are gaseous and in an active period of SF at these early times. However, in a relatively short
time, \mg/\ms\ strongly decreases in the inner 1 kpc region, excepting in runs Sp4S and Sp5S, which suffer late major mergers. 
The horizontal dotted line in the panels indicates
the value \mg/\ms=1/11. In the central 1 kpc regions, \mg/\ms\ falls below this value at redshifts from $\sim 4$ to $\sim 2$, 
evidencing that the inner regions start to 
quench early. At late epochs, $z\lesssim 0.5$, the values of \mg/\ms\ are $\sim 0.01$ or less; the innermost regions
of the simulated galaxies entered into a long-term quenching phase.  In contrast, the  global \mg/\ms\ ratios show a slow decreasing with
time and do not attain values below $\sim 0.1$ at $z\sim 0$, which suggests that the simulated galaxies are yet star-forming. 

In Figure \ref{sSFR}, the sSFR histories of the central 1 kpc (red line) and the whole galaxy (black line) are plotted.
Both the inner ($<1$ kpc) and global sSFR of the simulated galaxies are high at high redshifts: around 1Gyr$^{-1}$ or larger at $z> 3$, 
that is, both the global and inner stellar masses roughly duplicate by SF in $\lesssim 1$ Gyr. At lower redshifts, the sSFR strongly decreases. 
For the disk-dominated galaxies, the global sSFR decreases by roughly a factor of ten down to $z\sim 1$
and then continues decreasing but at a slower pace. At $z\sim 0$ these galaxies remain roughly as star forming, 
with values of sSFR around $2-5\times 10^{-2}$ Gyr$^{-1}$.
 The sSFR values inside 1 kpc are lower than the global ones almost at all epochs. 
The largest difference is for Sp8D: the inner sSFR falls strongly since $z\sim 1$ and at $z<0.5$ its value is on average 
$\lesssim 10^{-2}$ Gyr$^{-1}$ ($\mg/\ms\lesssim 5\times 10^{-3}$)
while the global sSFR is on average three times larger. The Sp8D galaxy presents the most prominent inside-out mass assembly trend among
all the simulated galaxies (see figures \ref{currentMGHs}--\ref{archMGHs}). Now we see that this is partially because 
the inner regions quench while the outer ones continue forming stars and growing in stellar mass. This implies that 
besides the structural inside-out mass growth, in our simulations there is an inside-out SF quenching process, which 
for Sp8D is very efficient.  For the spheroid-dominated galaxies, the inner and total sSFR are close at all redshifts. 

In Figure \ref{t-depletion}, the depletion time histories of the central 1 kpc (red line) and the whole galaxy (black line) are plotted.
Both the inner and global galaxy depletion times increase towards the present (excepting for Sp5S), showing that 
 the simulated galaxies were more efficient in transforming gas into stars in the past. 
In all the cases,  $\tdep$ in the inner 1 kpc is shorter than for the overall galaxy, that is, the inner galaxy regions are more efficient in
consuming gas into stars than the average of the whole galaxy. These differences tend to increase at lower redshifts. For the 
disk-dominated galaxies, at $z\sim 0.3-0$ the differences range from $\approx 0.4$ to 1 dex. The inner depletion times of these 
galaxies are mostly shorter than 1 Gyr, though in some short periods they may jump to higher values. With these short depletion times, 
if there is no gas replenishment, then the central galaxy regions quench relatively fast. In some cases and at high redshifts, 
the depletion time inside 1 kpc can be lower than 100 Myr, approaching to the typical timescales of the processes governing
the evolution of the interstellar gas.  Numerical simulations of gaseous galactic disks show that star-forming molecular clouds 
may form on timescales shorter than a few $10^7$ yr \citep{Dobbs+2012,Dobbs+2015}. Therefore, in the central 1kpc region of 
our galaxies and at high redshifts, the SF can be as efficient as within the environment of molecular clouds. 
For the spheroid-dominated galaxies, the differences in $\tdep$ between the inner regions 
and the overall galaxy are larger than for the disk-dominated galaxies, specially for lenticular-like galaxies Sp2L and Sp6L. 

The Sp2L galaxy shows a behavior different to the rest of the runs. The whole galaxy becomes very inefficient 
in consuming gas into stars (the global $\tdep$ increases with cosmic time up to $\approx 100$ Gyr at $z\sim 0$) in spite that the gas
is available; this galaxy ends with the highest gas-to-stellar mass ratio among all the simulations, \mg/\ms=0.35. 
However,  in its inner regions $\tdep$ remains with values around 1-2 Gyr since $z=0.5$, showing that the outer regions are those that
become with time very inefficient in transforming gas into stars; this ``outside-in'' behavior in the gas transformation efficiency into stars 
compensates the structural inside-out growth, and as the result, the radial stellar mass growth of this galaxy, especially the in-situ one, is nearly 
homogeneous and even with periods of outside-in assembly trends (see figures \ref{insituMGHs}, \ref{radial-distributions}, and \ref{age-profile}).

\section{Comparison with observational inferences}
\label{comparisons}

Our simulations of ``field'' MW-sized galaxies formed inside growing \lcdm\ halos make concrete predictions about how
the stellar mass is radially assembled across cosmic time. The half-mass radius grows fast with time and the radial 
stellar mass assembly is inside-out and driven mainly by in-situ SF/quenching. 
From the empirical point of view, there have been several attempts to infer how is the radius and radial stellar mass growth of 
galaxies, specially the MW-sized ones (see Introduction). Following we present some of these inferences and attempt to 
compare with our theoretical results. 
We calculate the average of the global and radial MGHs for two
types of MW-sized galaxies in our simulations: spirals (Sp1D, Sp3D, Sp7D, and Sp8D) + lenticulars (Sp2L and Sp6L), 
and highly spheroid dominated (Sp4S and Sp5S). The latter are spheroid dominated due to late major mergers, which produced 
also late bursts of SF (see figure \ref{SFRH}). Therefore, these galaxies can be associated to observed blue, 
star-forming early-type galaxies, which are 
rare in the local universe and are located in isolated environments \citep[see][and more references therein]{Lacerna+2016}.

\subsection{Fossil record inferences}
\label{fossil-record}

\citet{Ibarra-Medel+2016} have presented the normalized radial MGHs of the first release of MaNGA 
galaxies \citep{SDSS2016} using the fossil record method implemented in the specialized pipeline analysis
software called Pipe3D \citep{Sanchez+2016b,Sanchez+2016a}. Here we present the average normalized MGHs as in 
\citet{Ibarra-Medel+2016}\footnote{We use here a larger sample of MaNGA galaxies from the second data release \citep{Abolfathi+2017}
and with the updated data reduction pipeline \citep{Law+2016}, which improved the spectrophotometric calibration of galaxies.}
in the same stellar mass range of our simulations, $2-8\times 10^{10}$ \msun,\footnote{The stellar masses in Pipe3D 
are calculated by assuming a Salpeter initial mass function. We shift these masses by $-0.24$ dex to correct to 
a Chabrier \citep{Chabrier2003} initial mass function, which is close to the the \citet{MS79} one used in our simulations.}
and for two groups: spiral + lenticular galaxies (morphological type $T\ge 0$) to be compared with 
the average of runs Sp1D, Sp3D, Sp7D, Sp8D, Sp2L, and Sp6L, 
and blue, star-forming early-type ($T<0$) galaxies, to be compared with the average of runs Sp4S and Sp5S (see above).  
The morphological
classification of MaNGA galaxies has been performed by eye with some complementary criteria \citep[see][]{Ibarra-Medel+2016}, 
and the criteria of blue and star-forming galaxies are based on the color and sSFR distributions of SDSS 
galaxies as a function of \ms\ \citep[see][]{Lacerna+2014,Lacerna+2016}. 

As explained in \citet{Ibarra-Medel+2016}, in order to calculate
the average normalized MGHs, one needs to fix a given limit redshift, $z_{\rm lim}$, at which the final respective stellar mass is
defined for each galaxy. If $z_{\rm lim}$ is too low, then too few galaxies are left (the average redshift of the MaNGA galaxies
is around 0.035), and if $z_{\rm lim}$ is too high (in such a way that all the observed galaxies are included), then 
the late evolution (last $\sim 2 $Gyr for $z_{\rm lim}=0.15$) of most of observed galaxies is lost. As a compromise, 
we calculate the average normalized MGHs fixing $z_{\rm lim}$=0.08, which corresponds to 
 $\tlb\approx 1.0$ Gyr for the cosmology used here. The MGHs of galaxies observed at lower redshifts are interpolated and 
renormalized to start at  $z=0.08$.

\begin{figure}[htb!]
\plotone{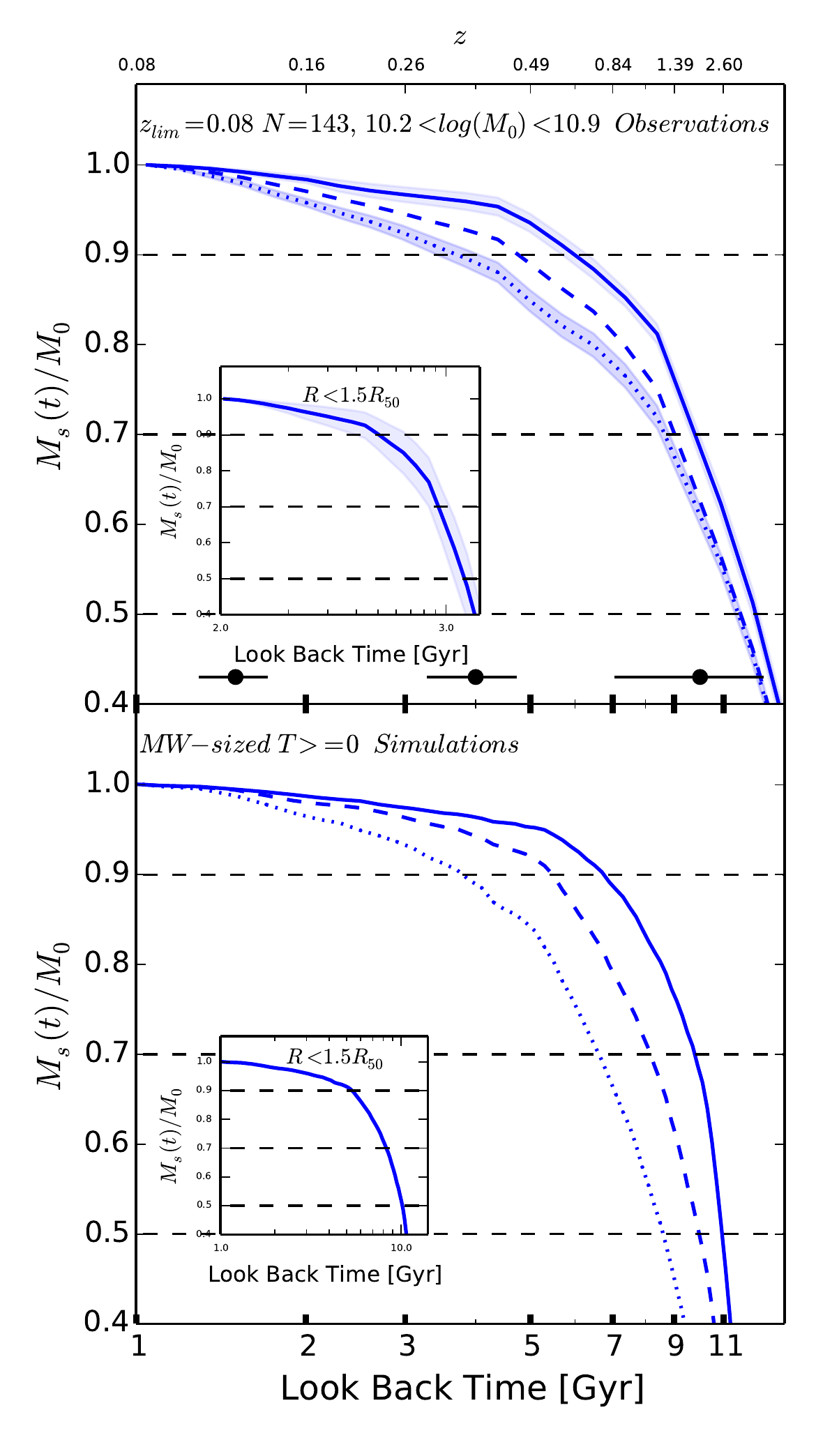} 
\caption{Normalized archaeological global and radial MGHs of disk/lenticular MW-sized galaxies from observations (upper panel) 
and simulations (lower panel). The global MGHs were calculated inside 1.5 effective radii, and the radial MGHs were calculated within radial bins 
corresponding to 0--0.5, 0.5--1, and 1--1.5 \reff; solid, dashed, and dotted lines, respectively. For the observations,
MW-sized galaxies of morphological type $T\ge 0$ and $i<75$ degree from the MaNGA/SDSS-IV sample (MPL-5 data
reduction pipeline) were used. The shaded areas are the errors of the mean. The horizontal error bars are estimates
of the methodological uncertainty in the determination of the stellar population ages at three look-back times.
For the simulations, the disk-dominated + lenticular-like galaxies were used. }
\label{MaNGA-MGHs}
\end{figure}

\begin{figure}
\plotone{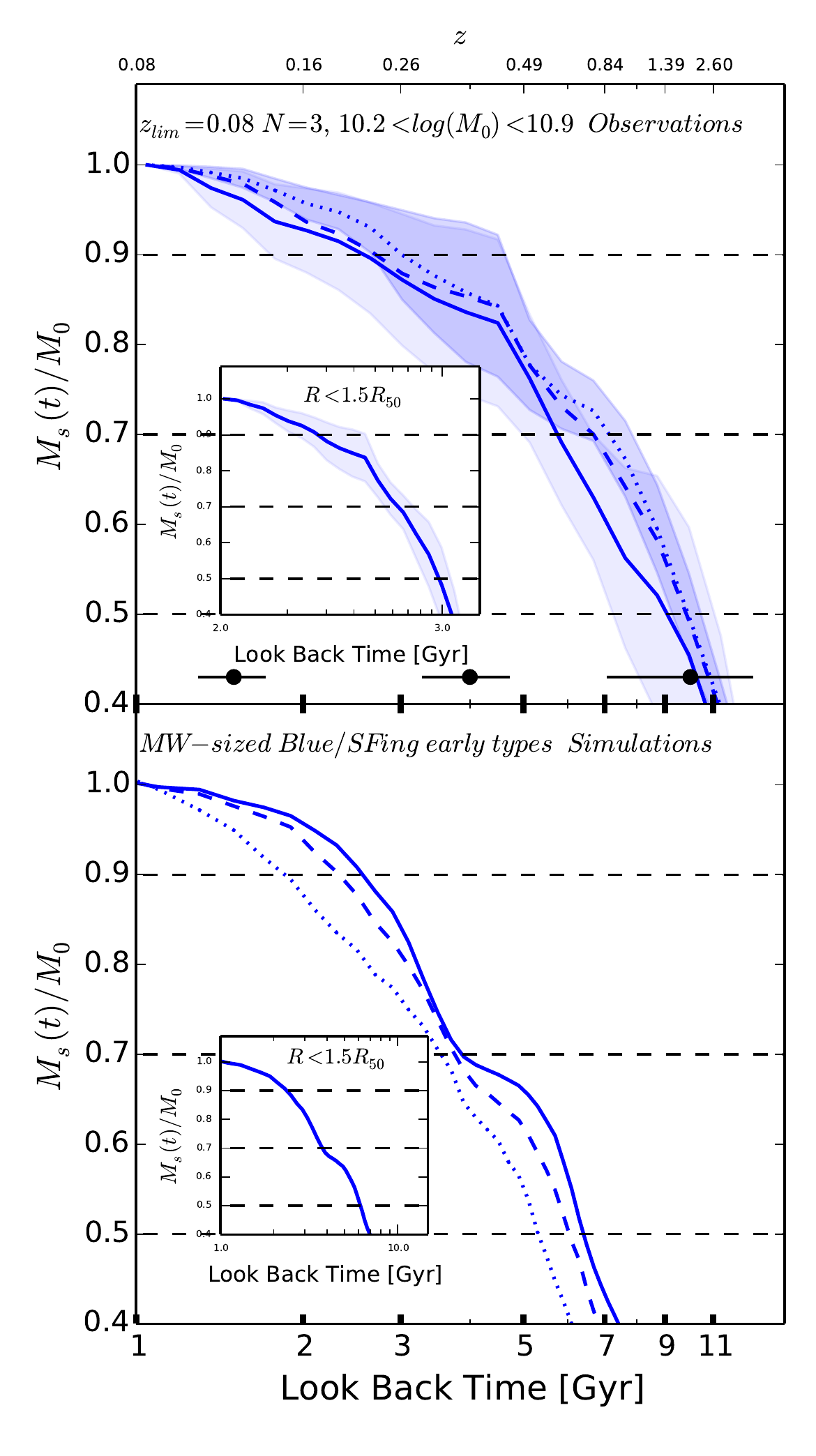} 
\caption{As in figure \ref{MaNGA-MGHs} but for MaNGA blue/star-forming early-type MW-sized galaxies (upper panel) and
the Sp4S/Sp5S simulated galaxies (lower panel). 
}
\label{MaNGA-MGHsb}
\end{figure}

The upper panels of figures \ref{MaNGA-MGHs} and \ref{MaNGA-MGHsb} show the average global and radial normalized MGHs of the 
MW-sized MaNGA subsamples of spiral/lenticular ($T\ge 0$) and  blue/star-forming early-type ($T<0$) galaxies, 
respectively, from $\tlb=1$ Gyr.
The shaded areas indicate the error of the mean for the global (inside 1.5 \reff; inset) and for the innermost and outermost radial MGHs
(solid and dotted lines, respectively). The radial MGHs correspond to the average in three radial bins: ($0,0.5$), ($0.5,1$) and ($1,1.5$) \reff,
where \reff\ is the effective radius of the galaxies in the $r$ band. The horizontal error bars give an estimate of the methodological 
uncertainty in the determination of the stellar population ages; the observed spectra poorly constrain the ages of the
oldest stellar populations.  The lower panels of  figures \ref{MaNGA-MGHs} and \ref{MaNGA-MGHsb} show the corresponding
average and radial normalized MGHs of our simulations, for which we use the archaeological radial MGHs (see subsection \ref{results}). 
The global MGHs (insets) are calculated within 1.5\re. The half-mass radius, \re, is typically smaller than the $r$ band effective
radius, \reff, but the differences are small \citep[on average 25\%\ in the $g$ band and less in the $r$ band,][]{Szomoru+2013}, 
so that the averages calculated in wide radial bins normalized to one or another radius are very similar. 

According to the insets of figure \ref{MaNGA-MGHs}, the average global archaeological MGH of $T\ge 0$ MW-sized galaxies from observations 
is slightly shifted at small mass fraction to larger look-back times with respect to the results from our simulations. The look-back times at 
which 50\% and 70\% of the final stellar masses are attained are on average 1.90 and 0.97 Gyr older for the observed galaxies than 
for the simulated ones, respectively; for the 90\% of the final mass, the average trend changes and the observed galaxies 
assembled 0.32 Gyr later than the observed ones.
As discussed in subsection 5.3 of \citet[][see also \citealp{Leitner2012}]{Ibarra-Medel+2016},  the statistical and systematic uncertainties in
the fossil record method seem to work in the direction of biasing the early mass assembly inferences to earlier epochs, and the late assembly to 
slightly later epochs.

The radial MGHs of the $T\ge 0$ galaxies show a clear inside-out trend, both for the fossil record inferences 
\citep[see also][]{Ibarra-Medel+2016} and the simulations. However, in more detail, {\it the simulations show  a more 
pronounced inside-out trend than the observational inferences.}  While the innermost normalized MGHs are quite similar, the
intermediate and outermost radial bins from the simulations assemble on average later than those inferred from observations. 
This difference can be partially accounted for the fact that most of galaxies are observed with some inclination with respect to the face-on position.  
For an inclined galaxy, the stellar populations along the line of sight of a given radial stellar bin are contaminated 
by the stellar populations of the other radial bins. As the result, the inferred radial MGHs tend to homogenize among them, 
the more as the inclination is larger. However, mock observations of some of our galaxies show that this effect becomes important
only for inclinations larger than $\sim 70$ degrees (Ibarra-Medel et al. 2017, in prep.). 

Regarding the rare blue, star-forming early-type galaxies in the $\ms= 2-8\times 10^{10}$ \msun\ range, there are only three in the
MaNGA current sample. The average normalized global MGH for them (inset in figure \ref{MaNGA-MGHsb}) reveals a late formation/assembly 
of their stellar populations (70\% of its mass formed at $\tlb\sim6.3$ Gyr). The simulated galaxies of 
this type also have a late stellar mass formation/assembly due to the late major mergers that they suffer, with 70\% of the mass  
formed on average at $\tlb\sim 4$ Gyr. For the 90\% of the assembled mass, $\tlb= 2.7$ and 2.4 Gyr for the average of the observations
and simulations, respectively.  The radial normalized MGHs, both for observations and simulations, are much closer among them
than in the case of the $T\ge 0$ galaxies. The effects of mixing stellar populations of different galaxies during the mergers
as well as merger-induced radial flows work in the direction of producing a nearly homogeneous radial distribution of 
stellar populations.  The observational inferences seem to suggest
even an outside-in formation, however, the error of the mean around each radial MGH is large. Looking individually each one of the three 
blue, star-forming early-type galaxies, two of them show periods of both inside-out and outside-in formation modes.  
We can conclude that at a qualitative level some observed blue, star-forming early-type galaxies can be explained 
as the product of late major mergers of disk galaxies, though the rejuvenation scenario (early formation, as most of
normal early-type galaxies, but late gas infall that triggers SF specially in the central regions) is also
possible (see subsection \ref{blueETGs}).

\subsection{Inferences from look-back time observations}

In \citet{vanDokkum+2013}, results of the inner and outer mass assembly of progenitors of MW-sized galaxies 
since $z\sim 2.5$ were presented. These authors have used the technique of linking progenitor and descendant 
galaxies by requiring that they have the same (cumulative) comoving number density
\citep[see e.g.,][]{Brown+2007,vanDokkum+2010}. The descendant galaxies at $z\sim 0$ are chosen to have 
stellar masses around $5\times 10^{10}$ \msun\
and for the selected progenitors, stacked images at different redshifts from the 3D-HST and CANDELS Treasury surveys are used. 
From the stacked mass surface density profiles of the progenitors, the evolution of the stellar mass contained inside and 
outside 2 kpc (physical scale), as well as the total mass, is presented.  

\begin{figure}[htb!]
\includegraphics[width=\columnwidth]{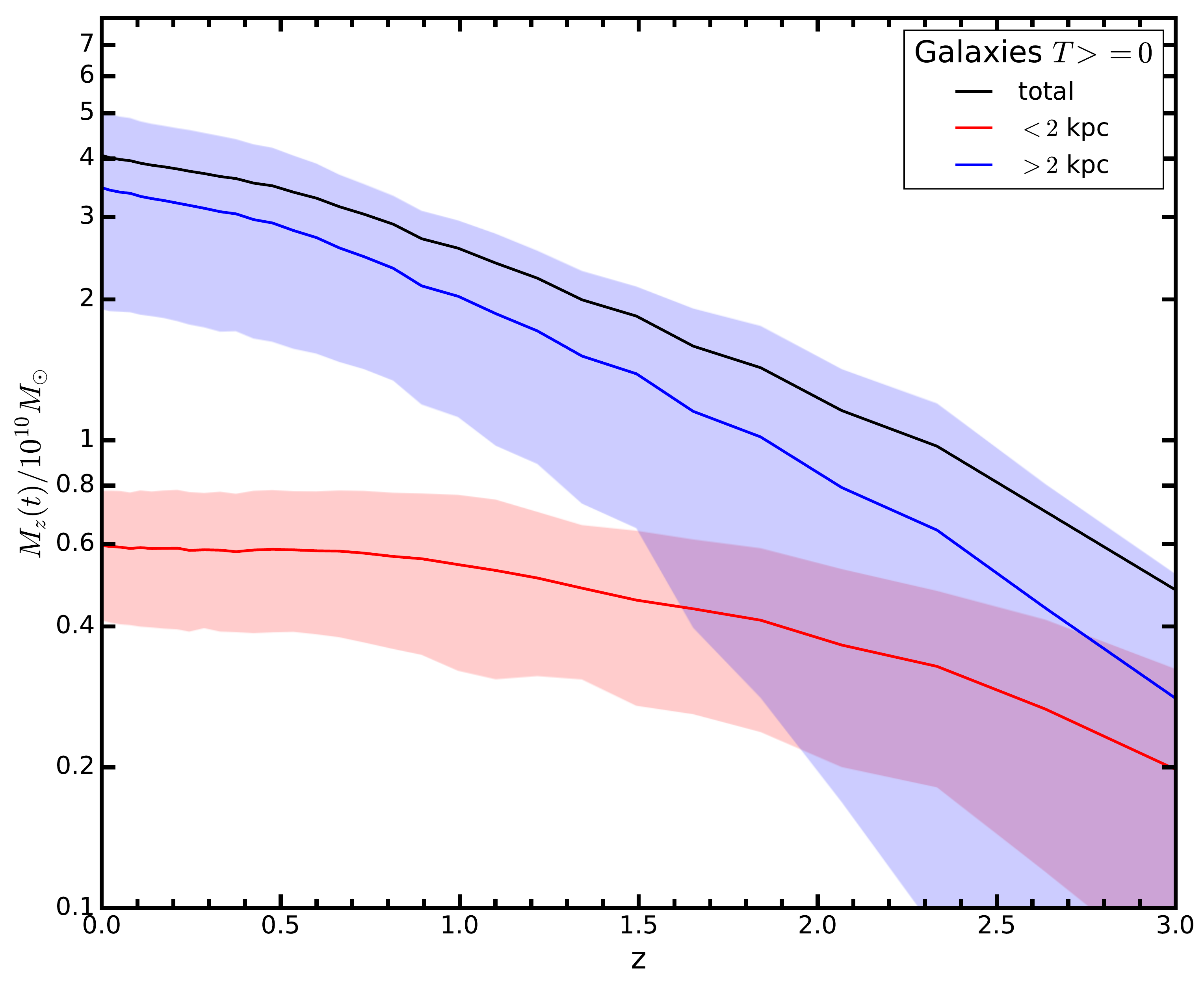}
\caption{Average global, inner,  and outer current stellar MGHs from the disk-dominated + lenticular-like simulated galaxies 
(late-type like, $T\ge 0$). Black line is for the total stellar mass, while red and blue lines are for the stellar mass inside 
and outside 2 kpc, respectively. The shaded areas 
show the standard deviations. This plot should be compared to the observational inferences by \citet{vanDokkum+2013} (their figure 4).
The outer regions grow on average much faster than the inner ones. The latter ones almost stop growing since $z\sim 1$; the 
SF quenches within the innermost 2 kpc. }
\label{vanDokkum-evol}
\end{figure}

Figure \ref{vanDokkum-evol} shows the average current MGHs inside and outside 2 kpc and the total MGHs, 
as in figure 4 of \citet{vanDokkum+2013} (red, blue, and black lines, respectively). For the averages and standard
deviations, we have used all the simulations but Sp4S and Sp5S since they correspond to very peculiar cases.
Simulations and observational inferences show the same trends: at high redshifts the mass within the inner 2 kpc grows with time 
slightly slower than the mass outer than 2 kpc, and at lower redshifts the difference between both rates increases to the point that the inner 
mass stops growing while the outer mass continues growing. In more detail, however, the simulations show an earlier total mass assembly 
and a larger difference between the inner and outer mass growth rates since $z\sim 1$ than the empirical inferences.
The strong slowdown of the inner mass growth starts at higher redshifts in the simulations ($z\sim 1$ on average) than in the
empirical inferences ($z\sim 0.6$). Note that the inner SF quenching in our simulations happens efficiently since high redshifts (see also
Section \ref{SFHs}), in spite of we are not including the effects of AGN feedback. 

Summarizing, {\it our simulations show a more pronounced inside-out stellar mass growth than 
the empirical inferences by  \citet{vanDokkum+2013} show. }

\begin{figure}
\includegraphics[width=\columnwidth]{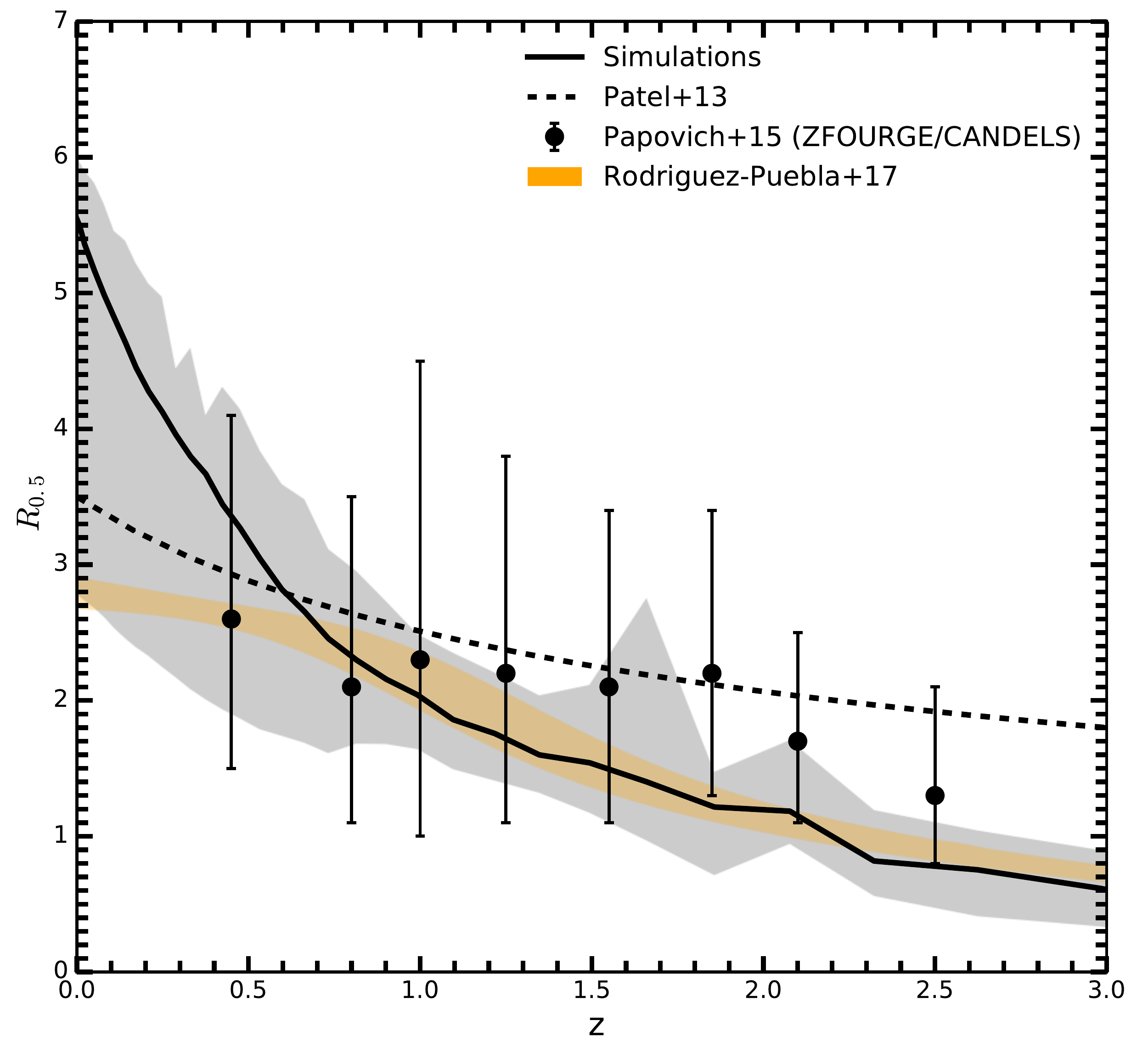}
\caption{ Median values of \re\ as a function of $z$ for the simulations (excluding runs Sp4S and Sp5S; solid line) and the percentiles
corresponding to a 1-$\sigma$ deviation (shaded area). The corresponding estimates from look-back observational studies
of MW-sized progenitors by \citet{Papovich+2015} and \citet{Patel+2013} are also plotted. The latter authors report only the 
best fit to their results. We also plot the  results from the semi-empirical approach by \citet{Rodriguez-Puebla+2017}, for galaxies
of present-day stellar masses between $2\times$ and $8\times 10^{10}$ \msun. }
\label{size-evol}
\end{figure}

\subsection{Half-mass radius evolution}

The empirical size evolution of MW-sized galaxies has been obtained by some authors by selecting the observed 
main progenitors at different redshifts
under the requirement that they have the same (cumulative) comoving number density at all redshifts \citep{vanDokkum+2013}
or by using the stellar mass growth inferred from the evolution of the star-forming sequence \citep{Patel+2013} or from abundance-matching 
techniques \citep{Papovich+2015}.  In Figure \ref{size-evol} we present the median half-mass radius evolution 
of our galaxies  (excluding runs Sp4S and Sp5S), as well as the percentiles corresponding to a $1\sigma$ deviation, 
and compare them with those obtained by \citet[][dots with error bars]{Papovich+2015} and 
\citet[][dot-dashed line for their fit to the data]{Patel+2013}. 
We also plot the semi-empirical inferences from \citet[][]{Rodriguez-Puebla+2017} for galaxies in halos with 
present-day masses between $0.8\times$ and $1.6\times 10^{12}$ \msun, corresponding to stellar masses
between $2\times$ and $8\times 10^{10}$ \msun\  (the orange shaded area encompasses this range of masses).
These authors, under some assumptions and using the observational constraints on the size--mass relation of star-forming
and quiescent galaxies at different redshifts, 
have extended the semi-empirical approach \citep[e.g.,][]{Behroozi+2013}
for constraining the radial stellar mass distribution of the evolving galaxies, assumed as disk-bulge systems
with disk and bulge mass fractions proportional to the fractions of  star-forming and quiescent galaxies, respectively. 

As seen in Figure \ref{size-evol}, the galaxy size increases as $z$ is lower both in simulations and in the empirical or 
semi-empirical inferences of MW-sized galaxies {\it but this growth is stronger in the former than in the latter}. This can 
be also seen from the evolutionary relation between \re($z$) and \ms($z$). According to \citet{vanDokkum+2013}, 
$\re(z)\propto \ms(z)^{\alpha}$, with $\alpha=0.27$ on average up to $z\sim 2.5$ (the average
slope is even shallower for the observations by \citealp{Papovich+2015}). 
For our simulations, very roughly $\alpha\sim 1$ at $z<1$ (the slope becomes flatter as $z$ is higher), i.e., the effective 
radius of  the simulated galaxies grows with mass faster than the observational inferences of \citet{vanDokkum+2013} 
and \citet{Papovich+2015}.   

For only the disk-dominated simulations and at $z\sim 0$, the median (mean) value of  $\alpha$ is 0.5 (0.7). That is, the 
growth in size is on average $\sim 0.5$ times the rate at which they grow in mass. This is larger than the value of 
$\alpha\sim 0.35$ determined by \citet{Pezzulli+2015} from combining empirical scaling relations of 
disk galaxies and assuming they do not evolve. These authors have tested that this value is consistent with 
observational determinations for a sample of late-type local galaxies using their stellar and SFR surface density profiles.

The strong \re\ growth of the simulated galaxies evidences a pronounced inside-out  structural growth and/or the effect of 
inside-out quenching of the SF. The latter is probably happening more efficiently in our simulations than in the observed galaxies.

\section{Discussion}
\label{discussion}

\subsection{Inside-out growth and inside-out quenching}

The rate at which increases the half-mass radius of our `field' MW-sized galaxies is on average faster than current observational inferences and 
simple predictions of disk galaxy evolution in growing CDM halos (see figures \ref{size-evol} and \ref{R-evolution}, respectively). 
These predictions are under the assumption of detailed 
angular momentum conservation and refer to the structural mass growth of the disks driven by the cosmological growth of the halos.  
For example, at $z=1$, for all the simulations but Sp2L, \re($z$=1)$=0.3\div 0.5$\re($z$=0),  while for the models by 
\citet[][see also e.g., \citealp{Somerville+2009}]{Firmani+2009}, \re($z$=1)$\ga 0.5\re$($z$=0). 
In the same way, the normalized radial current MGHs of our galaxies, specially  the disk-dominated ones,
 show a more pronounced inside-out stellar mass assembly than observational inferences (figures \ref{MaNGA-MGHs} 
 and \ref{vanDokkum-evol}). Following, we discuss on two physical processes present in the simulations that work in the 
 direction of amplifying the original structural inside-out growth of disks. 

{\bf Positive stellar feedback.-}
The adequate modeling of this feedback is a key ingredient to avoid the angular momentum catastrophe problem and to
form realistically looking MW-sized disk galaxies \citep[see e.g.,][]{Scannapieco+2008,Agertz+2011,Guedes+2011,Aumer+2013, Stinson+2013,
Marinacci+2014,Ubler+2014,Santi+2016, Colin+2016,Ma+2017,Grand+2017}. The feedback drives an efficient redistribution of the angular momentum 
of the baryons, which promotes the formation of extended disks, where stars form mostly in-situ out of the disk gas.  At a global level, the early 
conversion of low angular momentum gas into stars is lowered, making the SF histories flatter, with less SF at high redshifts 
and more SF at low redshifts \citep[][]{Ubler+2014}. A key feature of our simulated galaxies is that most of the stars are formed indeed 
in situ, and their SF histories are not strongly peaked at early epochs, showing even significant SF at late times \citep[see also][]{Grand+2017}. 
As shown in \citet[][]{Ubler+2014}, a   
fraction of the low-angular momentum ejected gas is re-accreted later (galactic fountain), but some of this re-accreted gas attains
higher specific angular momentum, eventually through mixing with the hot corona gas or from cosmic torques, and it infalls at larger radii.  
This process works in favor of strengthen the inside-out growth of the simulated galaxies.  As the result of the stellar-driven feedback, 
as shown in \citet[][]{Ubler+2014}, the specific angular momentum of the stars formed mostly in-situ out of the disk gas, becomes comparable or even
higher than the one of the halo. The specific angular momentum of our disk-dominated galaxies grows significantly with time, 
specially at late epochs \citep[see][]{Colin+2016}.  As in \citet[][]{Ubler+2014}, we find that the $z\sim0$ average specific angular momentum
of our disk-dominated galaxies is similar or larger than the one measured in their corresponding dark matter halos.  

{\bf  Inside-out SF quenching.-}
This is the second process that works in the direction of increasing the relative inside-out growth mode of galaxies.
The rate of mass growth due to SF suddenly changes from very fast to very slow in the innermost regions, while this  rate
changes much more gradually in the outer regions (excepting for the Sp2L galaxy; figure \ref{insituMGHs}). 
 Indeed, the inner regions (e.g., $< 1$ kpc) have typically much earlier SF histories than all the galaxy 
(figure \ref{SFRH}). The inner gas-to-stellar mass ratios, \mg/\ms=\tdep/\tsf, 
decrease to very low values since $z\sim 4-2$ (figure \ref{Rgas}). These values become much smaller than 0.1, suggesting that 
the central regions of the simulated galaxies enter into a long-term quenching phase very soon. Instead, the \tdep/\tsf\ ratios of the whole
galaxy do not decrease to values much smaller than 0.1. The metabolism of the inner regions of the simulated galaxies
is very different to the metabolism of the outer regions. 

The inside-out quenching seen in our MW-sized simulations, specially the disk dominated ones, strengthens the structural inside-out trend of 
the stellar mass assembly.  We highlight that this inside-out quenching process is without the presence of an AGN, and it is produced 
mainly by a burst of SF in the early compact galaxy, the consequent depletion of a high fraction of gas, the expulsion by 
SN-driven feedback of another fraction, and the absence of mechanisms of further significant gas replenishment to the central regions 
\citep[a qualitatively similar result has been found in numerical simulations at high redshifts by][see also \citealp{Zolotov+2015}, \citealp{Tacchella+2016b} ]{Tacchella+2016}. 
Once the disks are well stablished after $z\sim 1$, secular processes could contribute to transport fresh gas (and stars) 
to the centre, leading this to a more homogeneous radial stellar mass assembly. However, this seems not to be the case in our simulations, at least
not at a significant level. One of the main mechanisms of angular momentum and mass transport within the disk is the bar. In most of our
simulations bars are actually not formed efficiently (in a forthcoming paper, a detailed analysis on bar formation/destruction will be presented).

{\bf Other mechanisms.-}
\citet{Grand+2017} show that SF in their MW-sized simulations, as in our case, happens 
mostly in-situ, and with a radial distribution that may be categorized into either centrally concentrated or radially extended, 
inside-out formation.  For the simulations that end with large disk scalelenghts, they find that gas-rich quiescent mergers work as a 
mechanism by which the halo dark matter and gas acquire a high degree of specific angular momentum, which leads to 
high-angular momentum star-forming material condensing around the disk, and therefore to enhanced inside-out SF and large discs.
We will explore whether this mechanism is also relevant or not in our simulations elsewhere. \citet{Grand+2017}, 
who included AGN  feedback in their simulations,
show also that the effect of this feedback is mild in terms of the disk size. The most important effect 
of AGN feedback is to suppress central SF. This suppression prevents the formation of overly massive bulges in galaxies with 
high gas densities in the centre.

Finally, we note that the relatively fast evolution of the innermost galaxy regions in our simulations (early and strong SF peak and a consequent 
long-term quenching; figure \ref{SFRH}) implies that the simulated galaxies were too compact at high redshifts ($z> 2-3$). 
The half-mass radii a these redshifts are indeed small (figure \ref{size-evol}). A similar result has been reported by \citet{Joung+2009}, 
who claim that any potential viable solution to this apparent problem would have to reduce the amount of stars that are formed at very 
early epochs.  The authors suggest as possible mechanisms: (1) an early reionization with $z_{\rm ri}>>6$, a strong, internal
stellar- or AGN-driven feedback, or (3) a small-scale cut-off in the matter power spectrum, for instance if the dark matter is warm rather
than cold.

\subsection{The blue, star forming early-type galaxies}
\label{blueETGs}

In our suite of eight ``field'' MW-sized galaxies, two suffered late major mergers, making these galaxies to be spheroid dominated
but yet blue and star forming at $z=0$. In the local Universe these kind of galaxies are rare but they exist in isolated environments
\citep[e.g.,][and more references therein]{Lacerna+2016}.
Given the qualitative rough agreement found between the global and radial MGHs of simulated and observed galaxies
(subsection \ref{fossil-record}), we could conclude that the rare blue, star-forming early-type galaxies found in the local Universe can 
be explained as the result of late major mergers. However, there is room also for the mechanism of rejuvenation. In this
case, the blue, star-forming galaxy forms similarly to other early-type galaxies but lately suffers gas accretion, and 
the SF activates, specially in the center \citep[see for a discussion][]{Lacerna+2016}. In our small set of
simulations, we do not have this case.

\section{Summary and Conclusions}

We have studied the global and radial stellar mass growth histories, MGHs, of 
`field' MW-sized galaxies simulated with the N-body + Hydrodynamics ART code and presented previously in \citet{Colin+2016}. 
From our suite of eight simulations (see Table \ref{properties}), four are disk dominated at $z=0$ (D) 
two are lenticular like (L), 
and two are largely dominated by the spheroid but are blue and star-forming at $z=0$ (S; they suffered recent major mergers). 
Our main results and conclusions are as follow:

(i) The stellar half-mass radius, \re, increases with cosmic time, with some periods of shrinkage only at very early times or when
late major mergers happened in the case of runs Sp4S and Sp5S. The late growth of \re\ is very fast in all the runs, 
actually, faster than predictions of formation of disks in centrifugal equilibrium from baryons that conserve the 
angular momentum distribution of their growing \lcdm\ host halos (figure \ref{R-evolution}).  

(ii) The radial current normalized MGHs evidence an inside-out stellar mass growth, which is more pronounced in the disk-dominated
galaxies and less pronounced in the late-merging spheroid-dominated ones 
or even inverse at some epochs 
in the lenticular-like Sp2L galaxy (figure \ref{currentMGHs}). Since the current MGHs take into account gain/loss 
of stellar particles in/from the given radial bin as well as the stellar mass loss by winds, the radial MGHs some times can 
decrease. This happens specially when the galaxy suffers mergers. The in-situ radial normalized MGHs (accounted for only stellar particles 
formed in the given radial bin and ``frozen'' there, taking into account the mass loss by stellar winds) follow closely the respective 
current radial MGHs (figure \ref{insituMGHs}). This and other pieces of evidence show that (1) the radial stellar mass assembly 
of the simulated galaxies is driven by in-situ SF (the dynamical assembly due to mergers does not change significantly 
the inside-out growth mode), and (2) galaxies do not suffer a significant net radial stellar mass transport  in such a way  that 
the established radial stellar mass distribution does not change dramatically. 

(iii)  For the representative runs Sp8D and Sp6L, we have measured the radial displacement each stellar particle suffered.  
On average, indeed the particles do not show a net radial transport though they may shift in both 
radial directions by $\approx 1.8$ and 0.9 kpc at the 1-$\sigma$ level, respectively.  
We caution that due to the spatial resolution limit in our simulations (the cell length at the greatest level of 
refinement is $109 h^{-1}$ pc), dynamical processes associated to radial migration are likely underestimated.

(iv) The in-situ radial MGHs of all runs evidence an early fast growth by SF 
in the innermost radial bin followed by an abrupt slowdown, to the point that the innermost radial bin stops growing. 
The outer radial bins follow qualitatively this trend but much more gradual as more external they are. 
Within the inner 1 kpc, the sSFR decreases with cosmic time faster and the depletion time becomes much shorter than in all the galaxy,
specially for the disk-dominated simulations. As the result, the \tdep/\tsf\ ratios (equal to the gas-to-stellar mass ratios) inside 1 kpc
decrease much more abruptly than for all the galaxy. These ratios inside 1 kpc become $<<0.1$, showing that the inner
regions entered into a long-term quenching phase, while for the whole galaxy, these ratios remain above $\sim 0.1$. 
Therefore, our galaxies quench their SF the inside out, and as a consequence the mass growth by in-situ SF 
slowdowns more efficiently in the inner regions than in the outer ones.
The exception is run Sp2L (see above); this early-assembling lenticular-like galaxy is the one with the most
homogenous radial MGHs and highest gas fraction among all the runs. 

(v) The global and radial archaeological MGHs (constructed from the $z=0$ stellar particle age distributions) 
are similar to the current ones, showing this again that neither the mergers nor the radial net mass transport play 
a significant role in the structural evolution of our MW-sized galaxies. The exception is during the periods of 
major mergers. When galaxies suffered late major mergers, the archaeologically inferred MGHs are shifted to earlier 
epochs with respect to the current MGHs. 

(vi) We have compared our archaeological MGHs with those inferred by means of the fossil record method applied to the MaNGA survey 
\citep[][]{Ibarra-Medel+2016} in the same \ms\ range as our simulations. For the spiral/lenticular ($T\ge 0$) subsample, the predicted and observationally
inferred MGHs are in qualitative agreement, both evidencing the inside-out trend (figure \ref{MaNGA-MGHs}). However, the observational
inferences show a less pronounced inside-out trend than the simulations. 
For the rare blue, star-forming early-type galaxies 
(only two in our simulations and three in the observations), the MGHs are significantly later than those of the other galaxies, and qualitatively 
agree between simulations and observations (figure \ref{MaNGA-MGHsb}).

(vii) The average mass growth histories inside an outside 2 kpc agree qualitatively with empirical inferences from look-back observations 
of MW-sized galaxy progenitors \citep{vanDokkum+2013}: the outer regions grow faster than the inner ones (figure \ref{vanDokkum-evol}).  
However, this inside-out mode is more pronounced in the simulations than in the observations. The half-mass radius of the simulated galaxies also grows faster on average than estimates from look-back observations of the progenitors of MW-sized galaxies 
\citep{vanDokkum+2013, Patel+2013,Papovich+2015}. 

Since at MW-size scales the (uncertain) effects of stellar and AGN feedback are less critical for galaxy evolution than at other scales, it is
likely that our simulations do not differ substantially with respect to other recent simulations of MW-sized galaxies in spite of the
differences in the feedback implementations.
In this sense, we believe that the results presented here regarding the global and radial stellar mass assembly of
MW-sized galaxies formed in the \lcdm\ scenario are  generic, regardless of numerical details and feedback prescriptions.
Our main finding is that these galaxies assemble their masses the inside out, driven mainly by in-situ SF and by an inside-out process 
of SF quenching.  The latter happens without the presence of an AGN, in agreement with results found 
by \citet{Tacchella+2016} for high-redshift simulations. 

Qualitatively, our results are consistent with current observational inferences of the global and radial stellar mass 
growth from the fossil record method and from look-back time observations. However, in more detail, {\it the simulations
predict a more pronounced inside-out mode and a faster half-mass radius growth than all these observational inferences.} 
As discussed in Section \ref{discussion}, besides the structural inside-out mass build up, the simulated galaxies suffer 
also of some gas redistribution due to the SN-driven feedback (fountain effect) and of inside-out slowdown/quenching 
of SF (excepting Sp2L). Both effects tend to strengthen the inside-out mode and the rate of \re\ growth.
Are these effects too efficient in our simulations? Recall that we even did not include the effect of AGN-driven
feedback, which is expected to enhance the inside-out quenching of SF.   Do other simulations of 
MW-sized galaxies, where different schemes of SF and feedback were implemented, face the same potential problems
of too pronounced inside-out growth mode and half-mass radius growth? It is of great interest in the field to provide answers 
to these questions. So far, current numerical simulations and observational inferences start to be able to
constrain not only the evolution of the global properties of galaxies but also the local ones.

\acknowledgments

We thank an anonymous referee for useful comments and suggestions, which improved the quality of the manuscript.      
A.G-S. was financially supported by a UC-MEXUS Fellowship. H.I-M. was financially supported by a postdoctoral fellowship
provided by CONACyT grant (Ciencia B\'asica) 180125. 
This paper is dedicated to the memory of our friend and colleague Pedro Col\'in, who sadly passed away on January 14th 2017. 

\appendix

\section{Cumulative stellar mass distributions}
\label{mass-distribution}

\begin{figure*}[htb!]
\plotone{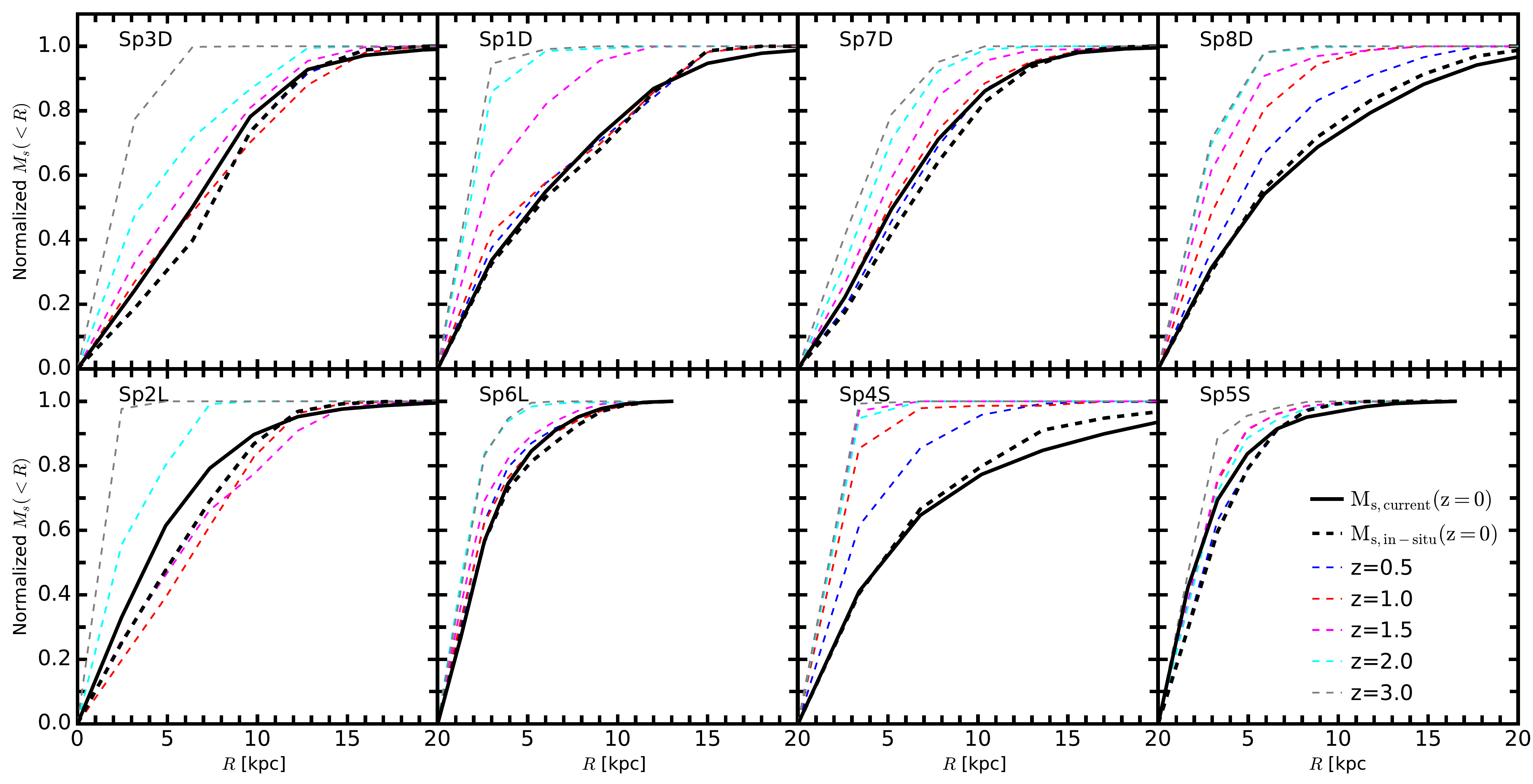} 
\caption{Cumulative stellar mass radial distribution, \ms($<R$), of the simulated galaxies at $z=0$ (solid lines). The black dashed lines are the 
cumulative radial distributions of the stellar particles formed in-situ at each radius and frozen there until $z=0$ (the stellar mass loss by winds 
is taken into account). If mergers and radial mass transport are not relevant in the structural mass assembly of the galaxy, 
then both cumulative radial distributions are expected to be similar. The in-situ frozen cumulative radial distributions are also shown at different redshifts  (color lines; see the color code in the inset). A clear inside-out growth of the stellar disks by in-situ SF is seen.}
\label{radial-distributions}
\end{figure*}

A  way to evaluate how much the radial stellar mass distribution of the galaxies has changed by net radial mass flows and/or 
mergers is by comparing the $z=0$ current cumulative radial stellar mass distribution, \ms($<R$), with the cumulative radial 
distribution of the stellar particles formed in-situ at each radial bin and frozen there, i.e., not taking into account the possible 
displacement of stellar particles from the given radial bin or particles that come from outside this bin (the stellar mass loss of 
the accumulated in-situ stars in the bin is taken into account), see also \citet{El-Badry+2016}.  
Such a comparison is shown in figure \ref{radial-distributions} for the eight simulated galaxies: black solid line is for the current 
radial mass distribution and black dotted line for the hypothetical in-situ SF radial mass distribution. In general, both 
distributions are very similar, showing that {\it the stellar particles formed in-situ did not suffer significant net radial re-distribution 
and that mergers do not alter too much the global radial stellar mass distribution. }
In four cases  the inner current cumulative radial mass distribution is slightly more concentrated than the one 
corresponding to the in-situ born stellar particles (the highest difference is for the Sp2L galaxy). This could be due to some external stellar 
mass incorporated into the inner regions (mergers), due to some inward stellar mass transport or due to outside-in SF quenching. 
In Section \ref{migration} we show that most of the radial stellar displacements are more radial mixing at relatively small scales 
than net inward or outward transport. 

In figure \ref{radial-distributions} are also shown the in-situ cumulative stellar mass profiles at redshifts higher than $z=0$ (color dotted lines). 
The evolution of the cumulative radial mass distribution of in-situ born stellar particles shows a clear inside-out growth in all the 
cases but the Sp2L galaxy (since $z\sim 1.5$). For the Sp2L galaxy, since $z\sim 1.5$ to $\sim 0.5$, the in-situ radial cumulative stellar 
mass distribution becomes slightly more compact. This is because the SFR slows down progressively from the outermost regions 
to the inner ones (see Section \ref{SFHs}).

\section{Radial displacements of the stellar particles}
\label{migration}

\begin{figure*}
  \centering
\includegraphics[width=7cm, height=5cm]{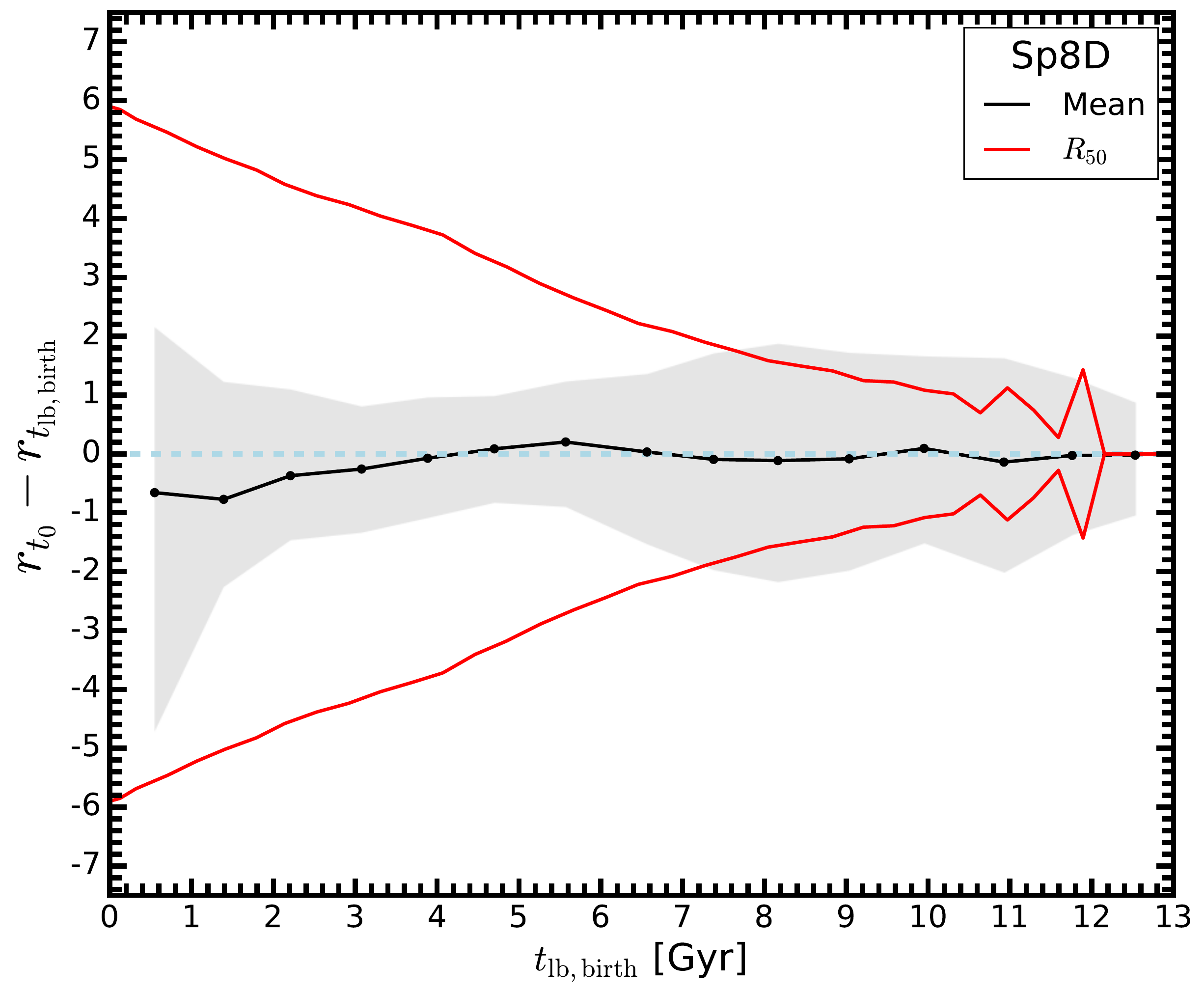}\includegraphics[width=7cm, height=5cm]{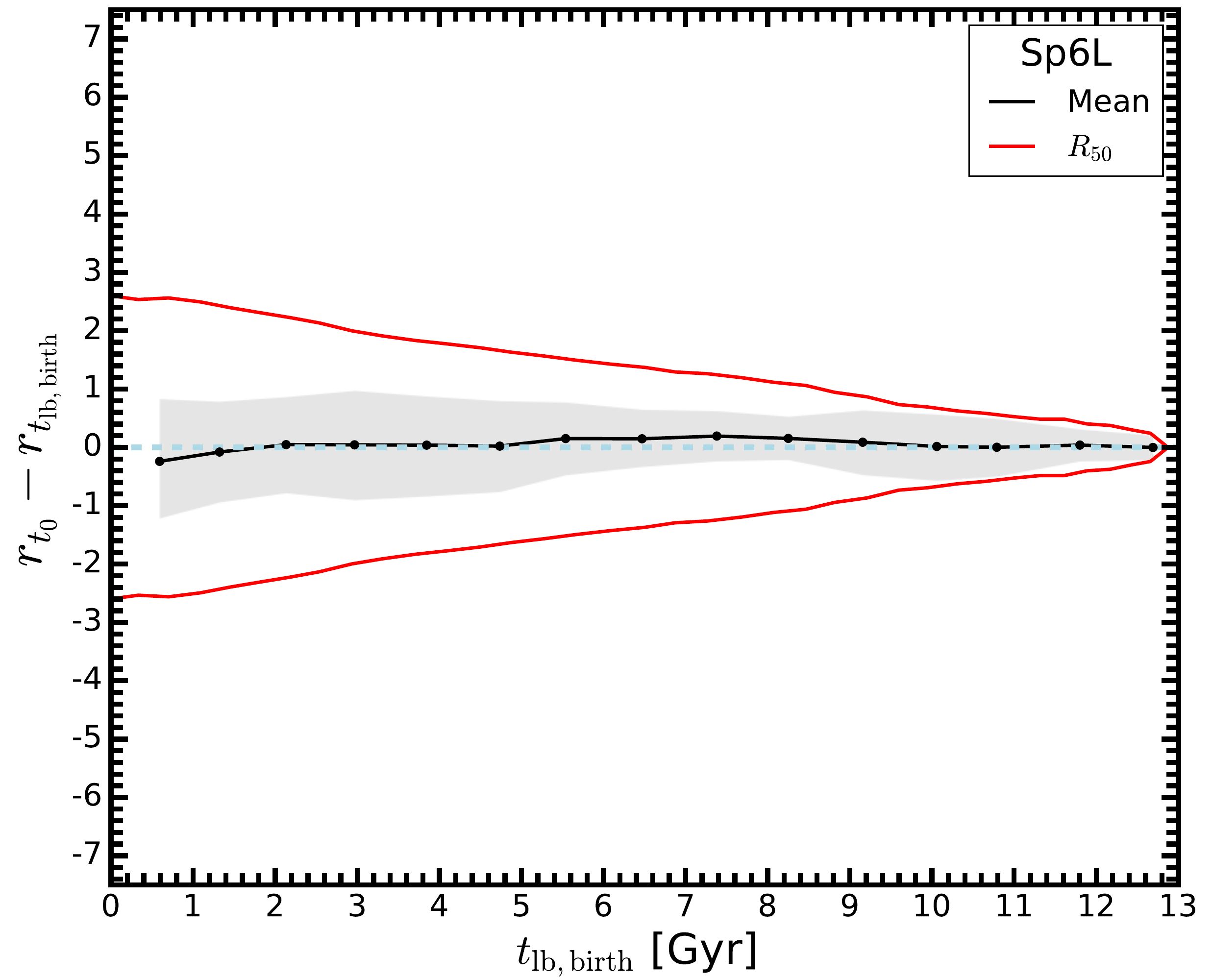}
\caption{Mean and standard deviation of the radial displacements suffered by the stellar particles since their birth at a given look-back time
and the present-day time for the runs Sp8D and Sp6L. The red line is the half-mass radius at each time both in the positive and negative 
side of the displacement.}
  \label{migr8-6}
\end{figure*}

The comparison between the current and in-situ radial MGHs presented in Section \ref{MGHs} and the comparison between the $z=0$ cumulative stellar mass profiles of all the stellar particles and those that formed only in-situ (figure \ref{radial-distributions}) strongly suggest that our simulated galaxies do not suffer significant radial mass redistributions by dynamical processes. Here, we measure directly the radial displacement of each stellar particle between its final position (at $z=0$) and  its position at birth, $t_{\rm lb,birth}$ for two representative runs, Sp8D and Sp6L.  
In figure  \ref{migr8-6} we present the results of this analysis, where the mean and standard deviation (solid line and shaded area) of the radial displacements in small bins of $t_{\rm lb,birth}$ are plotted.  
We also plot in these figures the half-mass radius at each time both in the positive and negative side of the displacement (red line)
with the aim to compare the radial shifts with the characteristic scales of the galaxy. 

The disk-dominated Sp8D galaxy presents a stellar MGH extended to late times and it is the one with the most pronounced 
inside-out growth (see figure \ref{currentMGHs}). Figure \ref{migr8-6} shows that the stellar particles on average remain close to the
radial positions where they were born, that is, there are not significant net radial mass flows. If any, there is a mild trend for the 
youngest particles, born mainly in the outermost regions, to
shift inwards on average by $\sim 0.5$ kpc (see also the pink line in figure \ref{currentMGHs}). This shift is very small compared to the 
characteristic scales of the galaxy at $z\sim 0$. In general, the stellar 
particles scatter similarly inward and outward with respect to their birth radius by no more than $\approx 1.8$ kpc at the 1-$\sigma$ level.
This kind of radial mixing is more pronounced for the particles older than $\sim 8$ Gyr; the 1-$\sigma$ shifts are actually larger than the
characteristic scales of the galaxy at the epochs these particles were born. Along with the net inward mass transport, there  
is also an increase in the radial mixing scales for the particles younger than $\sim 2$ Gyr, which are mostly the ourtermost particles. 
This could be related to the same dynamical process that produces warps in this simulation. 

The lenticular-like Sp6L galaxy presents an early stellar mass assembly, without mergers since early epochs and following an inside-out
growth mode (see figure \ref{currentMGHs}). Figure \ref{migr8-6} shows that the stellar particles on average remain close to the
radial positions where they were born, that is, again there are not significant net radial mass flows. The stellar 
particles scatter similarly inward and outward with respect to their birth radius by no more than $\approx 0.9$ kpc at the 1-$\sigma$ level.
These displacements are much smaller than the characteristic size of the galaxy since $\sim 9$ Gyr ago.

We should stress that the results presented above are very general. A more detailed analysis, taking into account the stellar particle
orbits, is necessary to study the question of stellar migration from the point of view of galactic dynamics.  
For this kind of studies, a higher spatial resolution than in our simulations is  likely required.
However, our numerical results seem to be in line with a radial mixing process happening in isolated MW-sized
galaxies rather than a net radial migration able to change significantly the stellar surface density profile 
\citep[e.g.,][]{Schonrich+2009,Roskar+2012}.

 \bibliographystyle{mn2efix}
\bibliography{references.bib}

\end{document}